\documentclass[aps,prb,english,twocolumn,longbibliography,
showpacs,floats, floatfix,10pt,superscriptaddress]{revtex4-2}
\usepackage{amsmath,bm,amsfonts}
\usepackage{mathtools}
\usepackage{romannum}
\usepackage{commath}
\usepackage{amssymb}
\usepackage{graphicx}
\usepackage{babel}
\usepackage{cases}
\usepackage[colorlinks=true, citecolor=blue, anchorcolor=green]{hyperref}
\hypersetup{linkcolor=red}
\usepackage{natbib}
\usepackage{floatrow}
\usepackage{xcolor}
\usepackage{braket}
\usepackage{physics}
\usepackage{array, appendix}

\newcolumntype{C}[1]{>{\centering\arraybackslash}m{#1}}

\newcommand{\Rmnum}[1]{\expandafter\@slowromancap\romannumeral #1@}
\newcommand{\ghz}{\text{GHZ}}

\newcommand{\avgc}{\langle g_3 \rangle}
\newcommand{\avgd}{\langle g_4 \rangle}
\newcommand{\avge}{\langle g_5 \rangle}

\newcommand{\avgtau}{\langle g_3(\tau) \rangle}

\newcommand{\be}{\begin{equation}}
\newcommand{\ee}{\end{equation}}
\newcommand{\bg}{\begin{aligned}}
\newcommand{\eg}{\end{aligned}}

\makeatother

\begin{document}
\title{Multipartite Greenberger-Horne-Zeilinger Entanglement
in Monitored Random Clifford Circuits}
\author{Guanglei Xu}
\affiliation{Institute of Physics, Chinese Academy of Sciences, Beijing 100190, China}
\altaffiliation{Present Affiliation: China Academy of Electronics and Information Technology, Beijing 100041, China}
\author{Yu-Xiang Zhang}
\email{iyxz@iphy.ac.cn}
\affiliation{Institute of Physics, Chinese Academy of Sciences, Beijing 100190, China}
\affiliation{School of Physical Sciences, University of Chinese Academy of Sciences, 
Beijing 100049, China}

\date{\today}

\begin{abstract}
Interactions in Many-body systems are typically short-range and few-body.
We investigate how such local interactions build up 
long-range and intrinsically multipartite entanglement 
by studying the $n$-partite Greenberger-Horne-Zeilinger ($\ghz_n$) entanglement 
in monitored random Clifford circuits, which is
well-known for a measurement-induced transition 
between phases of volume-law and area-law (bipartite) entanglement. 
We obtain a series of results: (1) About 1.25 $\ket{\ghz_3}$ can be 
extracted from states in the volume-law phase. 
This value is remarkably universal, 
independent of both the measurement rate and 
partitioning details, until a phase transition 
(either measurement-induced or a newly identified 
partitioning-induced transition) is approached.
(2) Dynamically, The creation (sometimes also the annihilation) 
of $\text{GHZ}_3$ entanglement occur suddenly via dynamical phase 
transitions (DPTs). The critical points of
these DPTs are governed by the entanglement speed ($v_E$) of biaprtite
entanglement. (3) In stark contrast to $\ghz_{n\leq 3}$, 
$\ghz_{n\geq 4}$ entanglement is statistically
significant only at the measurement-induced critical point, not in the 
bulk of the volume-law phase. 
Our results uncover a rich and previously overlooked hierarchy of 
multipartite entanglement structures.
\end{abstract}

\maketitle
\pagenumbering{arabic}

\section{Introduction}
Investigating many‑body physics 
through quantum entanglement has produced abundant landmark achievements
~\cite{Amico:2008aa,Eisert:2010aa,Laflorencie:2016aa,Abanin:2019aa}.
Nevertheless, entanglement is generally 
regarded as a comparatively coarse‑grained description of the 
full many‑body wave-function, because
entanglement is defined with respect to a partitioning of the system.
Correlations within the extensive 
internal degrees of freedom of each party are completely disregarded. 
This limitation also applies to the entanglement spectrum~\cite{Li:2008aa}, 
which, despite providing more detailed information, 
retains only inter‑party correlations.
To remedy this, a natural approach is to divide the system into more parties. 
Some recent works have made essential progresses
along this direction~\cite{Nezami:2020aa,Zou:2021aa,Siva:2022aa,Tam:2022aa,Liu:2024ab}. 

However, an issue that has not drawn enough attention yet is what physics that 
a measure of multipartite entanglement really quantifies, considering that patterns of correlation
could be extremely complicated in the multipartite scenario.
The measures chosen in some early attempts~\cite{Wei:2003aa,Wei:2005aa}
counts all possible entanglement between the subsystems. That is, 
supposing we divide the whole system into $n$ parties, such measures count 
all of $C_n^2=n(n-1)/2$ bipartite correlations, and $C_n^3$
tripartite correlations, and so on. In the end, one would like to have a multipartite
measure that capture the essential multipartite correlation that exists 
only when all of the $n$ parties present. Then, a hierarchy of $n$-partite entanglement
can be established by lifting $n$ from $n=3$. Indeed, the concept of
 ``genuine multipartite entanglement (GME)''~\cite{Horodecki:2009aa,Coffman:2000aa,Guhne:2009aa,Walter:2016aa,Sauerwein:2018aa,Bera:2020aa}
has been raised for this purpose. We will not repeat the definition of GME here but merely
emphasize that GME is proposed for practical settings of
multipartite quantum communication, which is not aligned well with 
quantum many-body systems. For example,
consider a tripartite state prepared by distributing many Bell pairs 
between every two of the three parties. This state perfectly fulfills the defining conditions of 
GME, although fundamentally it is only a collection of two-qubit Bell states.

This intuitive dissatisfaction with GME leads to the 
notion of \emph{irreducible multipartite entanglement} (IrME),
referring to GME that cannot be reduced to tensor products 
of entanglement belonging to a smaller group of parties.
Having established the concept of IrME, the subsequent questions are,
just like what people have asked about bipartite entanglement,
how to quantify IrME, how much IrME a quantum many-body 
system possesses, and 
how it arises during time evolution from a trivial product state.
We expect the answers to be significantly different from the case of bipartite entanglement,
especially for their emergence from dynamical evolution.
To see it, we observe that many-body systems typically 
have short-range and few-body interactions. Starting from a trivial product states,
such interactions immediately build up local structure of quantum entanglement,
which, however, does not contribute to IrME. 
Therefore, the dynamical features of IrME must be sharply different from
bipartite entanglement usually quantified by the so-called entanglement entropy.
Indeed, this observation motivated us to give a first systematic 
but still preliminary study of $n$-partite IrME.

To our best knowledge, there are only two measures of IrME
in the existing literature. One is the tripartite Markov gap~\cite{Hayden:2021aa},
which has been found to vanish for all ``triangle state''~\cite{Zou:2021aa}
hence meets the criterion of IrME. The content of tripartite Markov gap
has been calculated for the ground state of
conformal field theory~\cite{Zou:2021aa,Siva:2022aa} and found to be a
universal constant, $(c/3)\log 2$, where $c$ is the central charge. 
However, Markov gap is only defined for tripartite states. The other one is 
the measure of Greenberger-Horne-Zeilinger (GHZ) 
entanglement of an arbitrary $n$-partitite \emph{stabilizer states}~\cite{Fattal:2004aa,Bravyi:2006aa,Englbrecht:2022aa}. 
(A posteriori, generalizing to $n\geq 4$ is crucial, as it 
reveals fundamental differences from the case of $n\leq 3$.)
The $n$-qubit GHZ state
\be\label{ghz}
\ket{\ghz_n}=\frac{1}{\sqrt{2}}\big(\ket{0}^{\otimes n} +
\ket{1}^{\otimes n}\big)
\ee
has a distinguishing property that it is entangled as a whole, 
but any subsystem thereof is not.
It was proved in Ref.~\cite{Bravyi:2006aa} that 
by local unitaries in the form of 
$\otimes_\alpha U_{\alpha}$ where $\alpha$ labels the different parties,
any $n$-partite stabilizer state $\ket{\Psi_n}$
can be transformed into $g_n$ copies of $\ket{\ghz_n}$
and a residual state $\ket{\psi_n}$
\begin{equation}\label{decomposition-n}
\ket{\Psi_n}\xrightarrow{\otimes_\alpha U_{\alpha}} 
\ket{\ghz_n}^{\otimes g_n}\otimes\ket{\psi_n},
\end{equation}
where every copy of $\ket{\ghz_n}$ has exactly one qubit from
each party. Then, $g_n$ quantifies
$n$-partite GHZ ($\ghz_{n}$
in abbreviation) entanglement contained in $\ket{\Psi_n}$,
as long as no $\ghz_n$ content is in $\ket{\psi_n}$. 
The formula for $g_n$ will be introduced below. In the special case of $n=3$,
the residue state $\ket{\psi_3}$ can be further expanded 
into collections of Bell pairs and isolated qubits:
\be\label{eq:decomposition-3}
\bg
\ket{\psi_3}\rightarrow & \ket{\ghz_2}_{AB}^{\otimes n_{AB}}\otimes
\ket{\ghz_2}_{BC}^{\otimes n_{BC}}\otimes
\ket{\ghz_2}_{AC}^{\otimes n_{AC}} \\
&\otimes \ket{0}_A^{n_A}\otimes\ket{0}_B^{n_B}\otimes\ket{0}_C^{n_C},
\eg\ee
where  \textit{A}, \textit{B} and \textit{C} denote the three parties, and the notations such as $n_{AB}$ and
$n_A$, etc., denote the number of copies of each kind.

To have an idea about the content and dynamical feature of IrME, we thus
calculate $g_n$ in the \emph{monitored random Clifford circuits}~\cite{Li:2018aa,Chan:2019aa,Skinner:2019aa,Szyniszewski:2019aa,Gullans:2020ab,Fisher:2023aa}, 
which is a natural many-body playground for the dynamics of stabilizer states.
This system is well-known for the
measurement-induced transition between phases with
the volume- and area-law scaling of bipartite 
entanglement~\cite{Li:2018aa,Chan:2019aa,Skinner:2019aa,Szyniszewski:2019aa} and
an emergent statistical mechanics model~\cite{Bao:2020aa,Jian:2020aa,Zhou:2019aa}. 
The system has  been generalized to circuits with long-range
gates~\cite{Nahum:2021aa,Block:2022aa,Sharma:2022aa},
multi-qubit measurements~\cite{Lavasani:2021aa,Lavasani:2021ab},
spacetime duality~\cite{Ippoliti:2021aa,Lu:2021aa,Ippoliti:2022aa}
and qubits array of higher dimensions~\cite{Turkeshi:2020aa,Sierant:2022aa,
Liu:2022aa,Feng:2023aa,Jian:2021aa}. The literature also includes discussions of
error correction properties~\cite{Choi:2020aa,Gullans:2021aa,Fan:2021aa,Li:2021aa}, 
impact of noises~\cite{Weinstein:2022aa,Liu:2023aa,Liu:2024aa}, 
scrambling transition~\cite{Weinstein:2023aa}
and experiments~\cite{Hoke:2023aa}, etc.

In this Article, we shall report on 
three main results about $\ghz_n$ entanglement of the monitored 
random Clifford circuits:
\begin{enumerate}
\item States in the whole volume-law phase have a constant 
amount of $\ghz_3$ entanglement, $g_3\approx 1.25$, which is 
surprisingly insensitive to the size of the system ($N$), 
the rate of measurements ($p$), and the partitioning, in a large parameter regime.
A partitioning-induced phase transition is observed when one partition contains more
than $N/2$ qubits.\\
\item For dynamics, $\ghz_3$ entanglement does not grow up gradually but appears suddenly through
a dynamical phase transition (DPT)~\cite{Heyl:2018aa}, and sometimes dies 
through another DPT. We can determine the speed for the ``spread of
$\ghz_3$ entanglement'' through the dependence of the critical times of these DPTs 
on the partitioning. When there is no measurements $(p=0)$, this speed is found identical to
the \emph{entanglement speed} $v_E\approx 0.328$ of bipartite 
entanglement~\cite{Nahum:2017aa,Keyserlingk:2018aa,Nahum:2018aa}. \\
\item Steady-state $\ghz_{n\geq 4}$ entanglement 
is statistically significant only at 
the measurement-induced criticality. It is rare in both the volume- and the area-law phase. 
\end{enumerate}
Unfortunately, we do not have an analytical understanding to most of
our numerical discoveries. But we shall discuss our understanding of the relevance of $v_E$
to the dynamics of $\ghz_3$ entanglement, illuminating a competition between
bipartite and tripartite entanglement. We shall also discuss how the calculation
of $\ghz_n$ entanglement could shed light on general IrME.

This Article is arranged as following. In Sec.~\ref{sec:system}, we introduce the monitored
random Clifford circuits and the method for evaluate $g_n$.
In Sec.~\ref{sec:GHZ3}, we introduce the numerical results about the
steady state $\ghz_3$ content. In Sec.~\ref{sec:dynamics} we
study the evolution of $\ghz_3$ entanglement and introduce the dynamical
phase transitions of its birth and death. In Sec.~\ref{sec:GHZ45} we introduce
the results about the content of $\ghz_{n}$ entanglement
with $n\geq 4$. In Sec.~\ref{sec:conclusion} we conclude the paper and discuss what
the calculation of $\ghz$ entanglement may shed light on general IrME.

\section{Physical Setting and Preliminary}
\label{sec:system}
\begin{figure}[tb]
  \centering
  \includegraphics[width=0.97\textwidth]{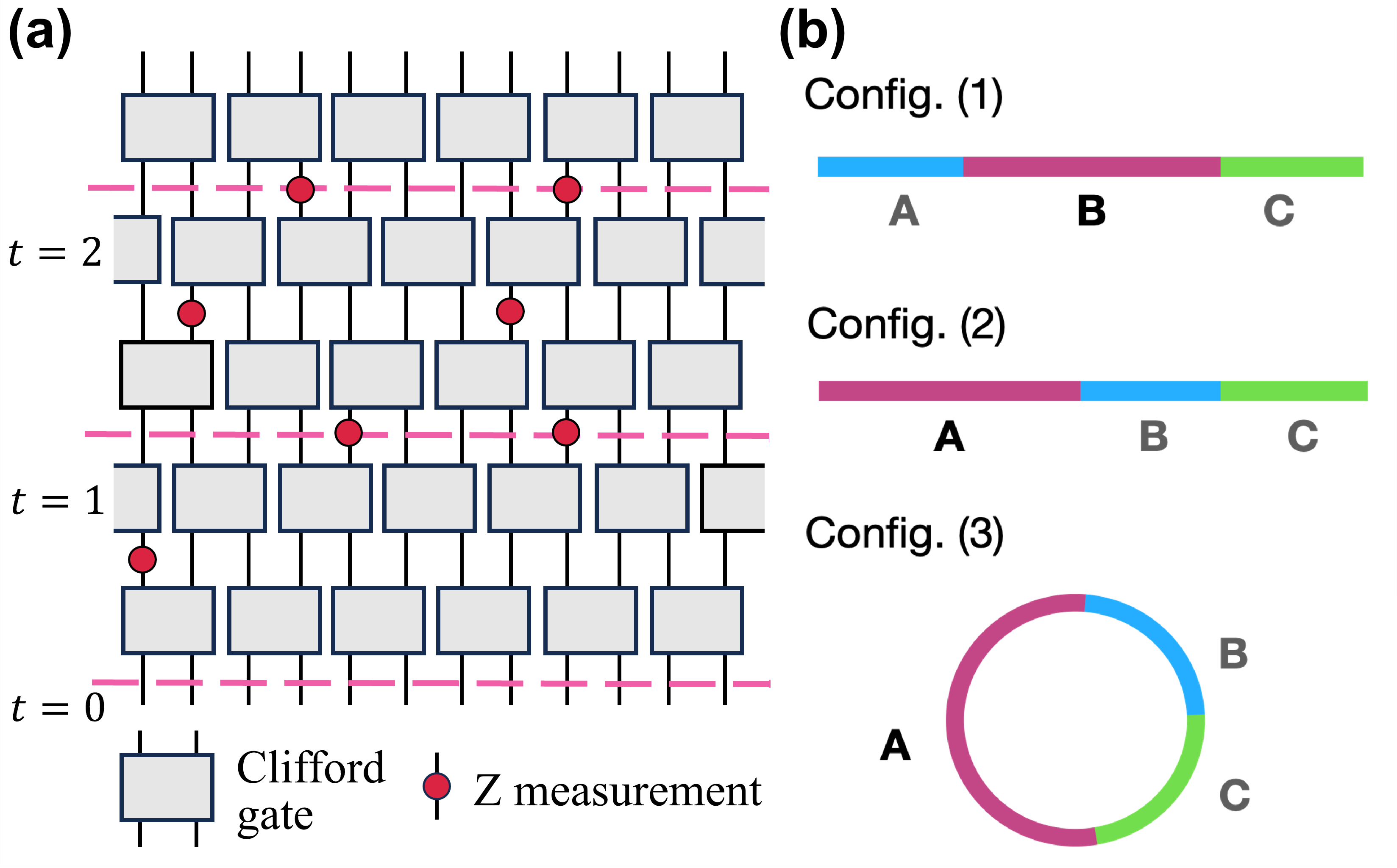} 
  \caption{(a) Setup: Random Clifford gates implemented
  upon neighboring qubits are arranged in a brickwork configuration.
  Each layer (counted by $t$) consists of two levels of random Clifford gates and
  after each level, projective Z-measurements are applied 
  independently to every qubit with probability $p$. (b) Illustration of the three configurations of
  tripartitions. In each of them, we vary the relative size of the marked party
  while keeping the other two equal.}\label{fig_circuit}
  \end{figure}

As illustrated in Fig.~\ref{fig_circuit}(a), we consider the prototype
brickwork circuits over a one-dimensional (1D) 
chain of $N$ qubits. Each rectangle represents a unitary randomly chosen from
the two-qubit Clifford group. After that, each qubit will be randomly measured along
the Pauli $Z$ basis with a probability of $p$. See also the caption of
Fig.~\ref{fig_circuit} for introductions of the circuit parameters. 
The state generated by Clifford circuits is called
\emph{stabilizer state}. A stabilizer state is a common eigenstate, with eigenvalue $+1$, of 
$N$ independent and mutually commuting operators called \emph{stabilizer generators},
or ``stabilizers'' in short. 
For example, we shall assume the 
initial state is $\ket{00\cdots 0}$ throughout this Article,
where $\ket{0}$ is the +1 eigenstate of the Pauli Z operators.
Hence, the initial state, although being trivial, is actually a stabilizer state
whose stabilizers are simply $\{ Z_1, Z_2, \cdots, Z_N\}$~\cite{Lavasani:2021aa}.
For stabilizer states, tracking the state evolution is equivalent to 
tracking these 
stabilizers in the Heisenberg picture~\cite{Gottesman:1998aa,Aaronson:2004aa,Anders:2006aa}.
These stabilizers are Pauli strings, i.e., products of
the identity operator $\mathbb{I}$ and
the standard Pauli operators $X, Y, Z$ of each qubit, 
with coefficients $\pm 1$ or $\pm i$. 
We say a stabilizer acts trivially on some qubits 
if the corresponding factor is $\mathbb{I}$. 

To evaluate $g_n$, we need to figure out 
$\mathcal{S}_{\text{loc}}$, the set of all elements 
of the stabilizer group that act trivially 
on at least one party. Denoting the number of its independent generators
by $\dim\,(\mathcal{S}_{\text{loc}})$,
it is proved that~\cite{Bravyi:2006aa}
\begin{equation}\label{eq:g}
  g_n=N-\mathrm{dim}\, (\mathcal{S}_{\text{loc}}).
\end{equation}
In our numerical calculations, $\dim\,(\mathcal{S}_{\text{loc}})$ is obtained from 
the binary representations of the stabilizers ~\cite{Nielsen:2010aa}
by Gaussian elimination, see Appendix \ref{sec:SM_S1} for details.
We notice that the decomposition~\eqref{eq:decomposition-3} allows us to 
evaluate $g_3$ by other means, e.g., 
$g_3=I_{AB}-2N_{AB}$~\cite{Sang:2021aa,Bertini:2022aa} where $I_{AB}$ and $N_{AB}$ denote
the quantum mutual information and quantum negativity between A and B, respectively. 
However, this does not help much for either numerical complexity or 
theoretical understanding. In the below we shall see that in the
volume-law phase $(p<0.16$) $g_3$ is an O(1) constant while both
$I_{AB}$ and $N_{AB}$ diverge with the increasing $N$. In this sense, evaluating
$g_3$ from their gap evokes the defining formula of the Euler constant.
Actually, there have been some theoretical analysis of the dynamics
of both $I_{AB}$ and $N_{AB}$ in the literature. Explicitly, the analysis based on
a mapping to the directed polymer in a random environment
has revealed the leading order ($\propto N$)
and the next-leading order of fluctuations ($\propto N^{1/3}$)~\cite{Li:2023aa,Weinstein:2022aa}. 
At these orders of approximation, one would obtain $I_{AB}=2N_{AB}$~\cite{Weinstein:2022aa},
which means $g_3=0$. This is simply because $I_{AB}$ and $N_{AB}$
are \emph{not} determined precisely enough: To get a reliable estimation of $g_3$, we
need precision at O(1) level.

\begin{figure*}
  \includegraphics[width=0.9\textwidth]{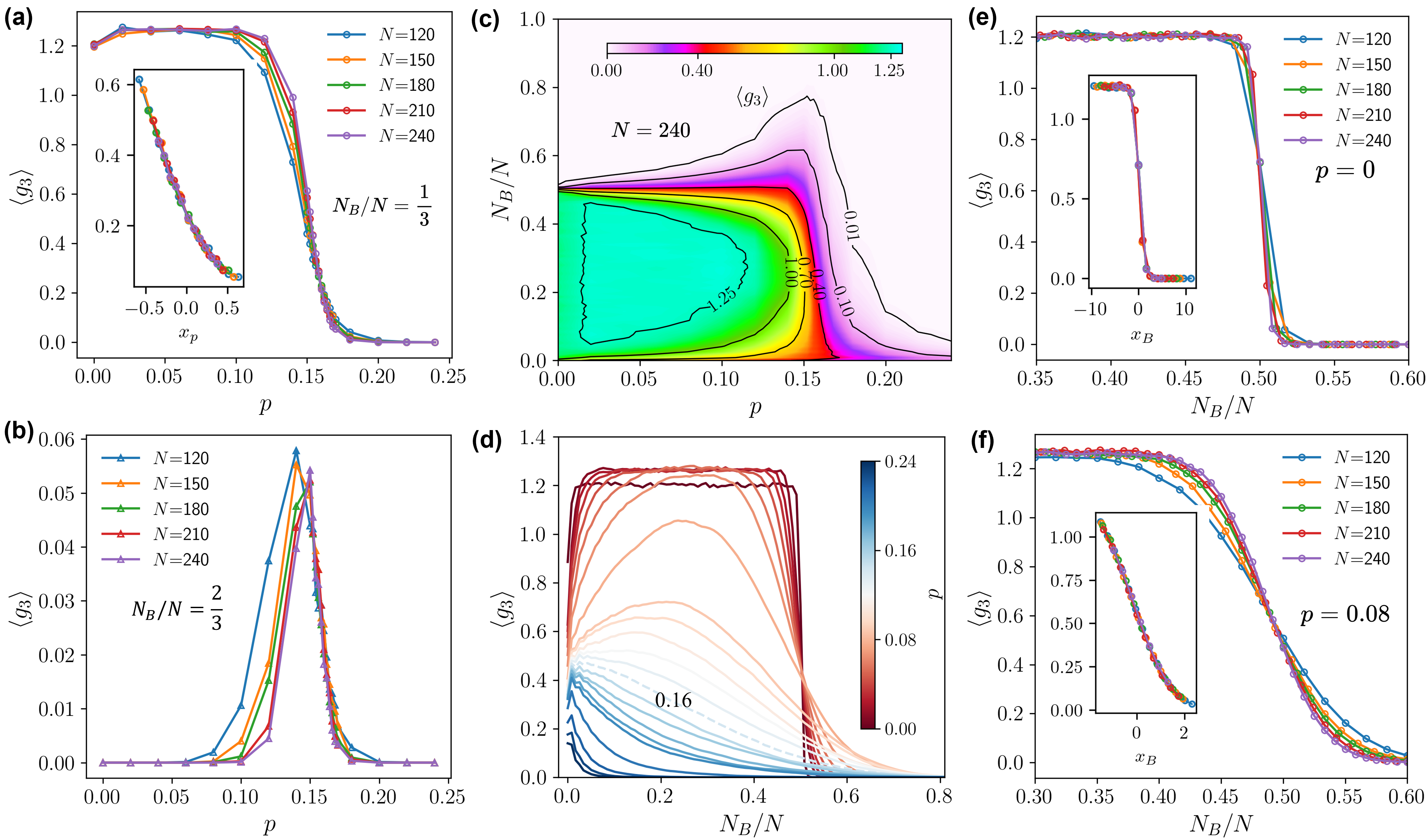}
 \caption{Measurement- and partitioning-induced $\ghz_3$ transition of circuits in Config.(1). (a,b) $\avgc$ as a function of $p$ 
with $N_B/N=1/3$ and $2/3$, respectively. The inset of (a) demonstrates the data collapse result, 
where the horizontal axis label is $x_p  \equiv (p-p_c)N^{1/\nu}$. (c) $\avgc$ as a function of 
  $p$ and $N_B/N$ for $N=240$. (d) $\avgc$ as a function of $N_B/N$ for
  $N=240$ and a series of $p$. (e,f) $\avgc$ as a function of $N_B/N$ for $p=0$ in (e) and 0.08 in (f). 
  The insets of (e,f) show the data collapse results,
  where the horizontal axis label is $x_B  \equiv (n_B- n^{cr}_B)N^{1/\mu}$ 
  with $n_B\equiv N_B/N$ and $n^{cr}_B=0.5$. The critical index 
  $\mu$ equals 0.989 and 1.541 for $p=0$ and 0.08, respectively.}  
  \label{fig_ghz}
\end{figure*}

In this Article, we shall consider tri-partitions in three 
different kinds of configurations illustrated in Fig.~\ref{fig_circuit}(b).
In Config.~(1), we assume open boundary condition for the circuit and 
Parties \textit{A} and \textit{C} have the
same number of qubits ($N_A=N_C$).
Then, the ratio between the size of \textit{B} ($N_B$) 
and the total number, $N_B/N$, parametrizes the partitioning. 
In Configs.~(2) and (3) with open and closed boundary 
conditions, respectively, we assume 
$N_B=N_C$ and parametrize the partitioning by $N_A/N$. 
The case of $N_A/N>1/2$ will be particularly interested. Our numerical
results shown in Fig.~\ref{fig_ghz}
and Figs.~(\ref{fig_transient},\ref{fig_ghz45}) are obtained by averaging over
2000 and 8000 samples of state trajectories, respectively. 
Every sampled circuit has $N$ layers (note that one layer has
two levels, so the circuit depth is $2N$). To evaluate 
steady-state $\ghz_n$, time average over the last 10\% layers is applied.

\section{Steady State $\ghz_3$ Entanglement} 
\label{sec:GHZ3}
In this section, we show the numerical results about the content of
$\ghz_3$ entanglement of the steady states. We put our focus on
circuits of Config.~(1), which has open boundary condition and the partitioning parameterized by
$N_B/N$. The case of Config.~(2) and (3) is summarized briefly in 
Sec.~\ref{sec:GHZ3_config23}.

\subsection{The Volume-Law Phase Has A Constant Amount of $\ghz_3$ Entanglement }
\label{sec:GHZ3-A}
Here the circuits are assumed to be of Config.~(1) by default. Our first main result is that
measurements also induce a phase transition of the
$\ghz_3$ entanglement. In Fig.~\ref{fig_ghz}(a),
we consider Config.~(1) with the equal partitioning $N_B/N=1/3$
and plot the averaged steady-states $\avgc$
as a function of $p$ for five values of $N$.
In the inset of Fig.~\ref{fig_ghz}(a),
we perform data collapse into the scaling form $\avgc =F[(p-p_c)N^{1/\nu}]$ with
$F$ an arbitrary function
and find that $p_c\approx 0.1596(4)$ and $\nu\approx 1.337(93)$. 
The critical value of $p$ is same with the
volume-law to area-law bipartite entanglement 
transition~\cite{Li:2018aa,Sierant:2022aa}, which is also 
interpreted as the purification transition~\cite{Gullans:2020aa}. 
There is no apparent connection between $\ghz_3$ entanglement and either 
bipartite entanglement or purification dynamics. Therefore, we propose that the 
$\ghz_3$ entanglement transition constitutes a third interpretation of the 
measurement-induced phase transition.

We observe a plateau of $\avgc\approx 1.25$ in Fig.~\ref{fig_ghz}(a), and
a small discrepancy at $p=0$ where $\avgc\approx 1.2$. 
This discrepancy is actually an odd-even effect. The results shown in
Fig.~\ref{fig_ghz} are obtained for even numbers of $N$, and even numbers of the
qubits in each partition.  We show in Appendix~\ref{app:odd-even} that if odd numbers are chosen, 
we will obtain $\avgc\approx 1.32$ instead. When $p\neq 0$, the qubits being measured
is removed from the entanglement, leaving the effective qubits number 
oscillate between even and odd, hence the absence of even-odd effect.
In Sec.~\ref{sec:dynamics-A} we shall meet another even-odd effect in the
dynamical evolution of $g_3$ depicted in Fig.~\ref{fig_transient}(a). 
Despite of such finite-size effect, we can claim that the volume-law phase is characterized
by a constant of 1.25 $\ghz_3$ entanglement.

Although a measurement-induced phase transition of $\ghz_3$ entanglement is not entirely unexpected,
we find it surprising and puzzling about the constant $\avgc\approx 1.25$ throughout the 
volume-law phase. This constant strongly suggests a hidden 
structure, which is protected against the random measurements.
Unfortunately, we do not have an insightful understanding of such structure. Nevertheless,
we present some discussions in the following. 

\subsubsection{A potential interpretation}
At first,  an upper bound of $\avgc$ can be derived in the case of $p=0$.
In Appendix~\ref{app:SM_S3} we use the fact that deep brickwork circuit
effectively generates random $N$-qubit Clifford unitary, and the decomposition~\eqref{eq:decomposition-3},
to derive an upper bound 
\be\label{eq:bound}
\avgc\leq \log_2 3.
\ee
This bound explains why $\avgc$ scales as O(1) in the volume-law, but still not sufficient to show 
why it is a constant independent to $p$. 

To have a partial understanding, we notice that Eq.~\eqref{eq:g} tells us that
$g_3$ equals the number of stabilizer generators supported on all the three parties.
In Config.~(1), such stabilizers are, 
up to multiplications by local stabilizers in $\mathcal{S}_{\text{loc}}$, long stabilizers 
penetrating the whole Party \textit{B}. One may relate such long stabilizers
to the \emph{logical qubits} of an error-correction code generated by 
the random Clifford circuits, hence understand its robustness to 
single-qubit measurements. The perspective of viewing monitored random circuits as a generator of
approximate error-correction code has been proposed in Refs.~\cite{Gullans:2020aa,Li:2021aa}.
However, these works, which described the qualitative properties of the volume-law phase, 
seems of limited use for the understanding 
of our quantitative  result of $\avgc\approx 1.25$. In the following, we show that
this constant also holds for a broad range of partitioning parameter.

\subsection{Partitioning-Induced Phase Transition}

Here we study the dependence of $\avgc$ on the size of each partition.
In Fig.~\ref{fig_ghz}(b), we tune the partitioning parameter to
$N_B/N=2/3$ and observe a different profile of 
$\langle g_3(p)\rangle$: The plateau disappears whereas 
a peak stands up near $p_c$. To have an idea of how it happens,
we plot $\avgc$ as a function of $N_B/N$ for $p=0$ and 
0.08 in Fig.~\ref{fig_ghz}(e) and (f), respectively. 
Interestingly, $\avgc$ is a constant until $N_B/N$ becomes larger
than 0.5. In the insets thereof, we show data collapse with 
the scaling form $\avgc =G[(N_B/N-1/2)N^{1/\mu_p}]$, where 
$G$ an arbitrary function. The data collapse confirms this as a phase transition.
We call it  \emph{partitioning-induced phase transition} (PIPT). 

To have a better idea of how $\avgc$ depends on $p$ and $N_B/N$,
we show a 2D plot for circuits of $N=240$ in Fig.~\ref{fig_ghz}(c).
The plateau of $\avgc\approx 1.25$ is demonstrated clearly. 
Unfortunately, we do not have a theoretical understanding of
such 2D plateau.  Next, we recast Fig.~\ref{fig_ghz}(c) into curves of $\avgc$ 
against $N_B/N$
for different values of $p$ in Fig.~\ref{fig_ghz}(d). 
PIPT is manifested by
a congruence of the red curves ($p<0.16$)
at $N_B/N=1/2$. The special role of $N/2$ is consistent with 
the numerical 
observation~\cite{Li:2019aa} that in the volume-law phase 
most of the stabilizers (in the clipped gauge~\cite{Nahum:2017aa}) 
have a contiguous length of $N/2$. This fact also helps to 
understand PIPT of circuits in Config.~(1):
when $N_B>N/2$, no stabilizer generators are long enough to 
cross all three parties, which is necessary for $\ghz_3$
entanglement according to Eq.~\eqref{eq:g}.

\subsection{The Case of Configs.~(2,3)}
\label{sec:GHZ3_config23}

We leave the detailed numerical results for circuits in Configs.~(2,3) in
Appendix~\ref{app:config23} but merely present the main difference from
Config.~(1) here.
Briefly speaking, the measurement-induced transition at $p=0.16$ 
still holds but PIPT is found
only at $p=0$. 
Recall that in Fig.~\ref{fig_circuit}(b) we label the
partitioning of Configs.~(2,3) by the ratio $N_A/N$. The absence of
PIPT at $p>0$ means that $\avgc\neq 0$ for $N_A/N>1/2$. 

In the above, we have attribute PIPT of Config.~(1) to the fact that no stabilizers
can be longer than $N/2$. This argument actually allows $\ghz_3$
entanglement in Configs.~(2,3) with $N_A/N>1/2$, 
because the three parties can always
be connected by a path shorter than $N/2$ despite of the size of \textit{A}.
On the other hand, we may argue that when $p=0$, the circuits are unitary and deep unitary circuits 
effectively sample the whole set of $N$-qubit Clifford unitary group, so that there will be no
difference between the three configurations.
In Appendix~\ref{app:SM_S3}, we show that if 
$N_A$, $N_B$, and $N_C$ do not respect the triangle inequality, the upper bound
in Eq.~\eqref{eq:bound} reduces to zero exponentially, 
confirming the PIPT at $N/2$ by statistical typicality.

How can the typicality argument be compatible with the argument on stabilizer length?
In the next section, we shall see that when $N_A/N>1/2$ 
the stabilizer-length argument really works, but only works temporarily:
We observe a transient $\ghz_3$ entanglement, which then
annihilates suddenly to restore the typicality prediction.

\begin{figure}[t]
  \includegraphics[width=1\textwidth]{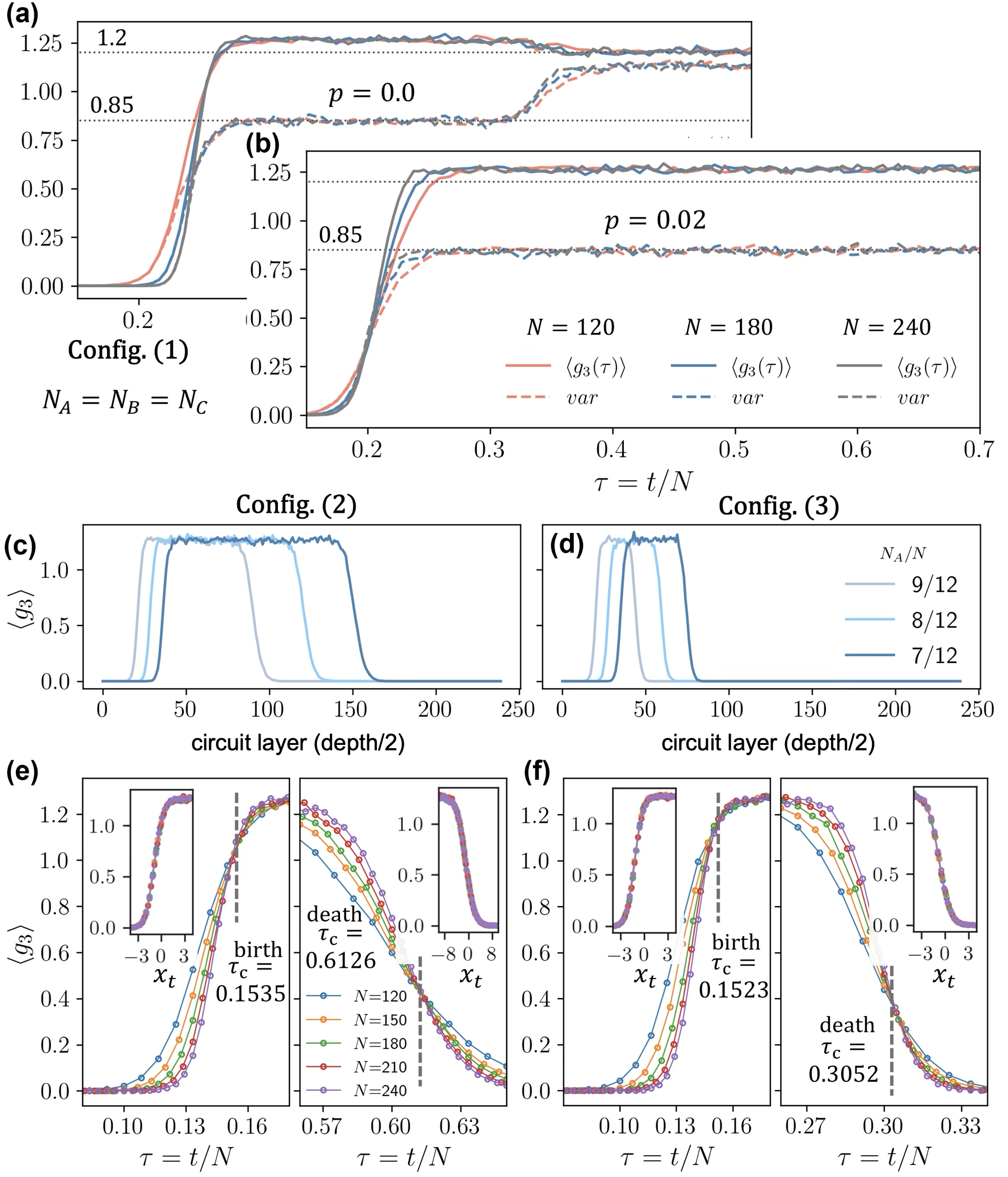}
 \caption{Dynamics of $\ghz_3$ entanglement. (a,b) Examples of
 $\avgc$ and the variances as a function of $\tau = t/N$  with $N=120, 180, 240$ 
 for the equal partitions ($N_B/N=1/3$), 
 and two different measurement probabilities $p=0, 0.02$. 
 (c,d) Examples of $\avgc$ as a function of circuit layer with $N=240$ 
 for three partitions of $N_A/N>1/2$ without measurements.
 (e,f) Dynamical phase transitions of the birth and death of $\ghz_3$
 in Configs.~(2,3), respectively, with $p=0$. }   
  \label{fig_transient}
\end{figure}

\section{Dynamics of $\ghz_3$ Entanglement} 
\label{sec:dynamics}

It is well known that bipartite entanglement grows linearly
in time after a quench. For $\ghz$ entanglement, however, we
have argued in Introduction that it is a non-local
correlation that cannot be built up directly by local interactions. Then, 
 there could be two possibilities for the growth of $g_3$:
\begin{enumerate}
\item Stays at zero for some duration, and then grows gradually until saturation at $\avgc\approx 1.25$; \\
\item Stays at zero for some duration, and then suddenly jumps to $\avgc\approx 1.25$.
\end{enumerate}
In the case of Config.~(1), the quantum mutual information between \textit{A} and \textit{C} actually builds up
in the first way (provided that $N_B<N/2$)~\cite{Nahum:2017aa}. 
We find that the dynamics of $\ghz_3$ follows the second way.
In this section, we will study the dynamics of $\ghz_3$ entanglement in the 
case of $p=0$ detailedly. The case of $p>0$ is briefly discussed.

\subsection{The Birth and Death of $\ghz_3$ Entanglement via Dynamical
Phase Transitions}
\label{sec:dynamics-A}

Firstly, we revisit the
equal partitioning of Config.~(1) and plot
the ensemble average (solid curves) and the variance (dashed curves) 
of $g_3$ for
${p=0}$ and 0.02 in Fig.~\ref{fig_transient}(a) and (b), respectively. 
Therein, the horizontal axis is normalized layer
$\tau=t/N$ [recall that $t$ equals the doubled 
circuit depth, cf. Fig.~\ref{fig_circuit}(a)]. 
For $p=0$, the three stages of $\avgtau$ have
heights 0, 1.25, and 1.2, sequentially. The three-stage-profile 
is also seen in the variance (0, 0.85, 1.1, sequentially). 
As mentioned above,
the 3rd stage is where the odd-even parity appears. 
In the curves of $p=0.02$, we see only two stages,
where the 2nd
stage has the same $\avgc$ (1.25) and 
variance (0.85) with the intermediate stage of $p=0$.
Then, we return to the cases of
Configs.~(2) and (3) with $N_A/N>1/2$ and $p=0$.
Surprisingly, as shown in 
Figs.~\ref{fig_transient}(c,d), we observe
transient $\ghz_3$ entanglement 
in the 2nd stage, where the
plateau height is also $\avgc\approx 1.25$.

The birth and death of $\ghz_3$ entanglement 
have characters of DPTs. 
We choose Configs.~(2) with $p=0$ and 
$N_A/N=3/5$ and plot $\avgtau$ for five different values of $N$
in the left panel of Fig.~\ref{fig_transient}(e),
where the inset shows the data collapses into an arbitrary function
$F[x_t]$ with $x_t=(\tau-\tau_c)N^{1/\mu}$.
The death of the transient $\ghz_3$ is found to be another DPT, 
as evidenced by the right panel of Fig.~\ref{fig_transient}(e).
DPTs for the birth and death of $\ghz_3$ are also found
in Config.~(3), see Fig.~\ref{fig_transient}(f).
The normalized critical depth (doubled $\tau_c$) are
\begin{subequations} \label{eq:DPT}
\be
d_{\text{birth}}  = \frac{l_B}{2v_E},
\ee
\be
d_{\text{death}}  =\begin{cases}
  2l_B/{v_E} &\text{Config.~(2)}  \\
  l_B/{v_E}  & \text{Config.~(3)}
\end{cases}
\ee
\end{subequations} 
where $l_B=N_B/N$ and $v_E\approx 0.328$ is the 
\emph{entanglement speed}
defined by the averaged increment of bipartite entanglement per 
circuit level~\cite{Keyserlingk:2018aa}.
The formula of $d_{\text{birth}}$ applies to all three configurations with $p=0$. 
Unfortunately, we do not have an analytical derivation of
these DPTs from the first principle. Nevertheless, we present our unsophisticated 
understanding of the critical times as follows.

\subsection{Critical Time of The Death-DPTs}

We start from death-DPTs of $\ghz_3$ entanglement 
and discuss why it must dies when $N_A/N>1/2$ at the above critical times.  
The central idea is to transform into an ``interaction picture'',
which manifests the competing roles played by the bipartite entanglement.

\begin{figure}[t]
  \centering
   \includegraphics[width=0.85\textwidth]{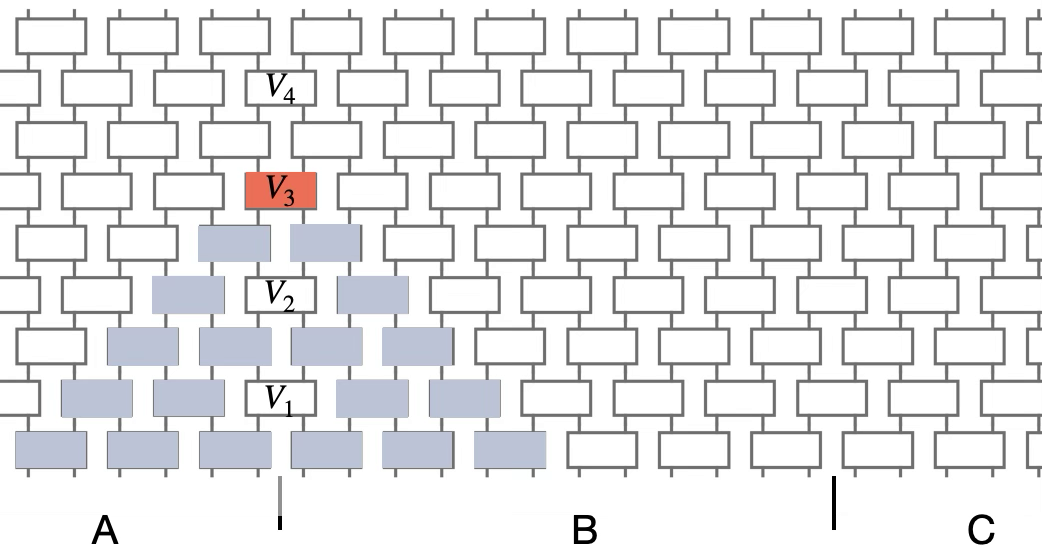} 
  \caption{An illustration of gates that contribute 
  to $\tilde{V}_k$ with $k=3$. Now \textit{BC} is viewed as a single party
  so that the interaction occurs only at the boarder between \textit{A} and \textit{B}.
  } \label{fig_em}
  \end{figure}

Let us start from Config.~(2) and denote the untiary realized by a circuit 
of depth $d=2t$ by $U_{ABC}(t)$. We can view 
$U_{ABC}(t)$ as the evolution unitary
in the conventional Schr\"{o}dinger picture. An interaction picture 
can be introduced by referring the
``free evolution'' to the gates 
acting within either \textit{A} or the joint system of \textit{BC}.
The ``interaction'' part is thus the gates crossing the boarder between
\textit{A} and \textit{B}. Then, the unitary evolution in 
this interaction picture is 
\be \label{em:inter-picture}
\tilde{U}_{\textit{A-BC}}(t)
=U_A^\dagger(t) U_{BC}^\dagger(t) U_{ABC}(t)
\ee
A short derivation shows that 
\be
\tilde{U}_{\textit{A-BC}}(t)=
\tilde{V}_t\tilde{V}_{t-1}\cdots \tilde{V}_1
\ee
where $\tilde{V}_k$ is associated to the $k$-th gate coupling 
\textit{A} and \textit{BC}, $V_k$, via the formula
\be
\tilde{V}_{k}\equiv U_A^{\dagger}(k)U_{BC}^{\dagger}(k) V_{k}
U_A(k)U_{BC}(k).
\ee
Gates that contribute to $\tilde{V}_k$ are marked in Fig.~\ref{fig_em}.
Similarly, we introduce the interaction picture within
the subsystem of \textit{B} and \textit{C}:
 $\tilde{U}_{\textit{B-C}}$ as 
\be 
\tilde{U}_{\textit{B-C}}(t)=U_B^\dagger(t)U_C^\dagger(t)U_{BC}(t).
\ee
The Schr\"{o}dinger picture is thus related to the 
interaction picture via  
\be \bg 
U_{ABC}(t) & =U_{A}(t)U_{B}(t)U_C(t)
\tilde{U}_{\textit{B-C}}(t) \tilde{U}_{\textit{A-BC}}(t) \\
& \sim \tilde{U}_{\textit{B-C}}(t) \tilde{U}_{\textit{A-BC}}(t)
\eg\ee 
where $U_{A}(t)U_{B}(t)U_C(t)$ is omitted 
in the second line because it is irrelevant to entanglement. 
Thus, $U_{ABC}(t)$ is equivalent to firstly applying $\tilde{U}_{\textit{A-BC}}$, 
and secondly $\tilde{U}_{\textit{B-C}}$.
And in each step, the system is treated as bipartite.

The effect of $\tilde{U}_{\textit{A-BC}}$ is to generate state
\be
\ket{\Psi'(t)}_{\textit{A-BC}}=\tilde{U}_{\textit{A-BC}}(t)\ket{0,0,\cdots 0}.
\ee 
If $N_A>N_B+N_C$ (hence $N_A/N>1/2$), \textit{BC} will be fully entangled
with \textit{A} at the circuit depth
\be 
d=(N_B+N_C)/v_E+O(N^{1/3}),
\ee 
where the second describes the fluctuation of random circuits. 
In the large $N$ limit and the condition that $N_B=N_C$,
we obtain the normalized critical depth $d_{\text{death}}= 2l_B/v_E$.
The fact that \textit{BC} is fully entangled with \textit{A} means that
the joint state of \textit{BC} is maximally mixed:
\be \label{eq:mixed_BC}
\rho'_{\textit{BC}}=(\frac{\mathbb{I}}{2})^{\otimes (N_B+N_C)}. 
\ee 

The next step is to consider $\tilde{U}_{\textit{B-C}}(t)\ket{\Psi'(t)}_{A-BC}$. 
The unitary works on \textit{BC} but now $\rho'_{\textit{BC}}$ is the maximally mixed
state, hence, $\tilde{U}_{\textit{B-C}}(t)$ cannot bring any correlation between \textit{B} and
\textit{C}. Indeed, forming $\ghz_3$ entanglement implies an entropy of \textit{BC} 
lower than that of state~\eqref{eq:mixed_BC}.
Therefore,  $\ghz_3$ entanglement is ruled out.

The same reasoning applies also to Config.~(3), where the closed boundary 
condition is used. Due to the closed boundary condition, $\textit{A}$ contacts with both \textit{B} and \textit{C}.
Thus, von Neumman entropy of \textit{BC} increases
with the speed of $2v_E$, leading to $d_{\text{death}}= 2l_B/v_E$.

\subsubsection{The Absence of Death-DPT when $p>0$}

Above we have seen that the fact crucial to the death of
transient $\ghz_3$ entanglement is state~\eqref{eq:mixed_BC}. 
A maximally mixed state is only possible if both
$N_A/N>1/2$ and $p=0$ are fulfilled. Now suppose we turn
the measurements on. Since a measurement 
resets a qubit to pure state either $\ket{0}$
or $\ket{1}$, a state trajectory of \textit{BC} will never be maximally mixed,
so that we cannot expect a death-DPT based on the above argument.

\begin{figure}[tb]
    \centering
    \includegraphics[width=0.8\textwidth]{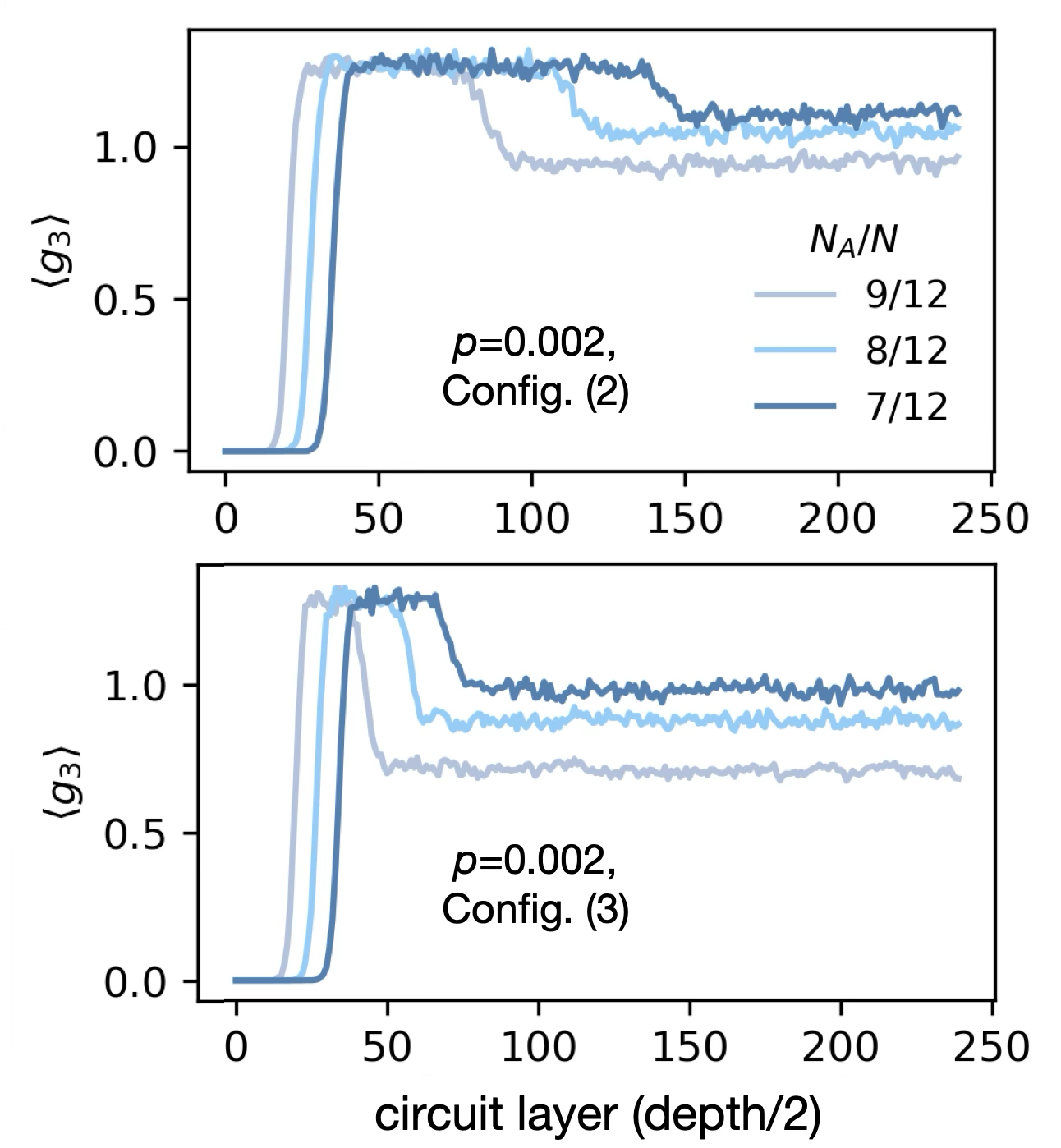}
    \caption{ The 3rd stage of the time-evolution of $\ghz_3$ entanglement behaves like a lower plateau when $p$ is 
    0.002 and $N=240$. The other parameters are same with Figs.~\ref{fig_transient}(c) and (d).  }
    \label{fig:partial}
\end{figure}

It is interesting that our finite-size numerical calculation
captures an intermediate regime, where the transient $\ghz_3$ entanglement
becomes less but not completely disappear. This occurs for $p< 0.01$ when $N=240$.
We show some curves in Fig.~\ref{fig:partial}, where the upper and lower
panels display the cases of Configs.~(2) and (3), respectively. 
In the plots, we can see a three-stage profile where the 3rd plateau is lower than 1.25 but 
higher than zero, demonstrating a ``partial death'' 
of the $\ghz_3$ entanglement
We consider such partial death of the transient $\ghz_3$ entanglement as a finite-size effect,
which occurs if $p$ and $N$ satisfy some relation.

\subsection{Critical Time of The Birth-DPTs} 

At first, we notice that conditions for the emergence of $\ghz_3$ has been briefly discussed in 
Ref.~\cite{Bertini:2022aa}. It was found that
$\ghz_3$ is impossible for causality reasons, if the circuit depth is shallower 
than $N_B/(2v_{\max})$, where $v_{\max}$ is identified as 
the light-cone speed $v_{\max}=1$. The causality bound applies to every individual state trajectory whereas
here we focus on statistical properties. Our numerical results improve the bound from
the light-cone speed  $v_{\max}=1$ to the bipartite entanglement velocity
$v_{E}\approx 0.328$. It means that although 
being allowed by causality, $\ghz_3$ is still statistically 
negligible unless the circuits are deep enough.

Next, we give an argument for the relevance of $v_E$ in the 
case of $p=0$. The argument is based on the observation that
random circuits tend to maximize the von Neumman 
entropy of each subsystem (thermalization at infinite temperature). 

The argument goes as follows. For tripartite stabilizer states, 
entanglement is contributed by either 
$\ghz_3$ entanglement or the Bell entanglement.
At the critical point of the birth-DPT, $d_{\text{birth}}=l_B/(2v_E)$,
\textit{B} has obtained the maximal entropy.
Since then, entropy increase can only come from 
\textit{A} and \textit{C}. We notice that
forming \textit{A-C} Bell entanglement and forming $\ghz_3$ entanglement 
(for example, by developing an \textit{A-B} Bell 
entanglement to the tripartite $\ghz$)
contribute equally to the entropy increase 
of subsystem either \textit{A} or \textit{C} individually.
But only forming $\ghz_3$ entanglement helps to 
increase the entropy of the joint system \textit{AC}. 
Thus,  $\ghz_3$ entanglement becomes statistically significant.

To compare, no such advantage of $\ghz_3$ exists before 
\textit{B} is maximally mixed (or perhaps more precisely, before the entropy of \textit{B} is saturated). 
In this case, new Bell pairs can then 
form between \textit{A-B}, \textit{B-C}, and \textit{A-C}.
The averaged contribution to von Neumann entropy
of these pairs is equal with $\ghz_3$. Therefore, we concludes that
$l_B/(2v_E)$, the time when \textit{B} becomes the maximally mixed,
is the critical time of birth-DPT.

\subsubsection{Measurements Move Birth-DPTs Forward}
\label{sec:dynamics_p}

\begin{figure}[tb]
    \centering
    \includegraphics[width=0.85\textwidth]{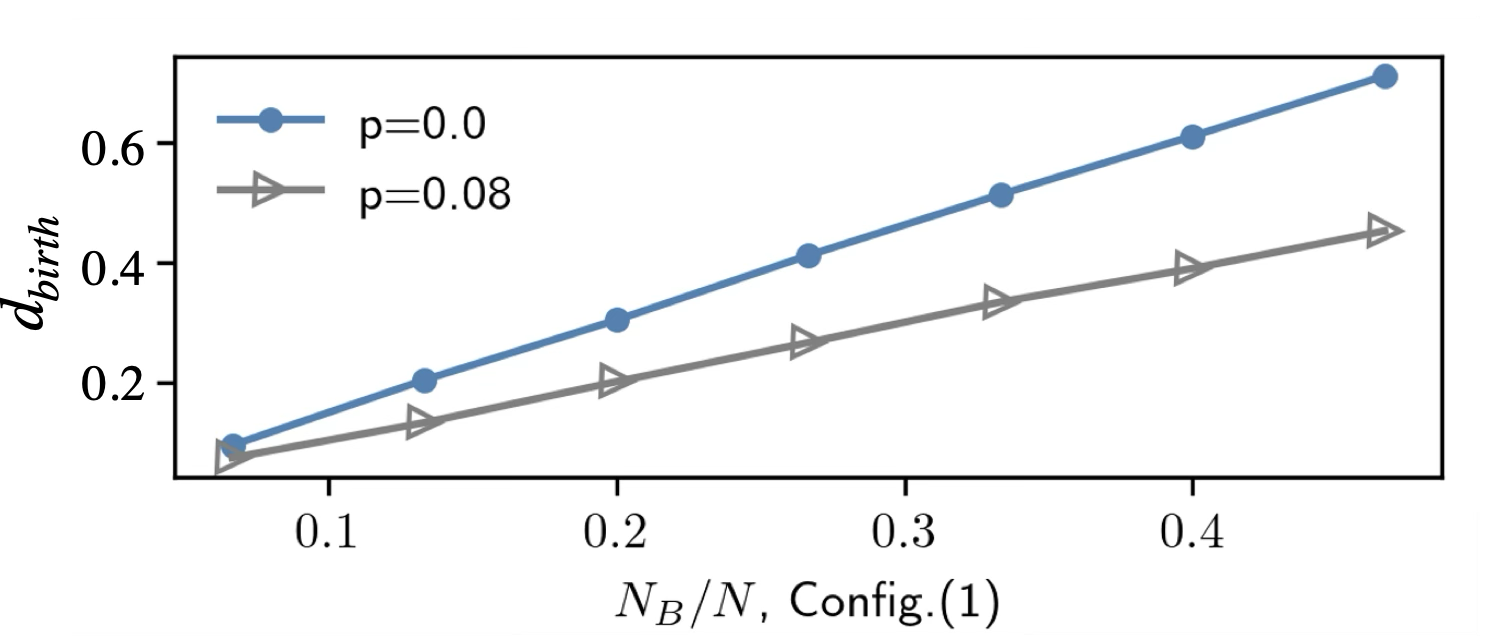}
    \caption{The critical points of the birth-DPTs of 
    $\ghz_3$ entanglement as
 functions of partition parameter in Configs.~(1) for two values of $p$.}
    \label{fig:lifetime123}
\end{figure}

Our numerical calculation shows that $d_{\text{birth}}$ becomes smaller when
$p>0$, see Fig.~\ref{fig:lifetime123} for an example. 
In this sense, we say that measurements move the birth-DPTs forward. 
Meanwhile, measurements reduce bipartite entanglement so that
$v_E$ decreases by its definition. 
Thus, Eq.~\eqref{eq:DPT} becomes surely invalid. 

This effect is of course related to the phenomenon of measurement-induced long-range
entanglement. Nevertheless, our above argument suggests that $\ghz_3$ entanglement appears 
when the entropy of \textit{B} is ``saturated.'' When measurements
are turned on, a portion of the qubits of  \textit{B} are thus always
dis-entangled from the others. Thus, entropy of \textit{B}
saturates at a lower level and within a shorter duration. This explains
why birht-DPTs come earlier when $p>0$.

\section{$\ghz_{n\geq 4}$ Entanglement as A Critical Phenomenon} 
\label{sec:GHZ45}

\begin{figure}[tb]
  \centering
   \includegraphics[width=\textwidth]{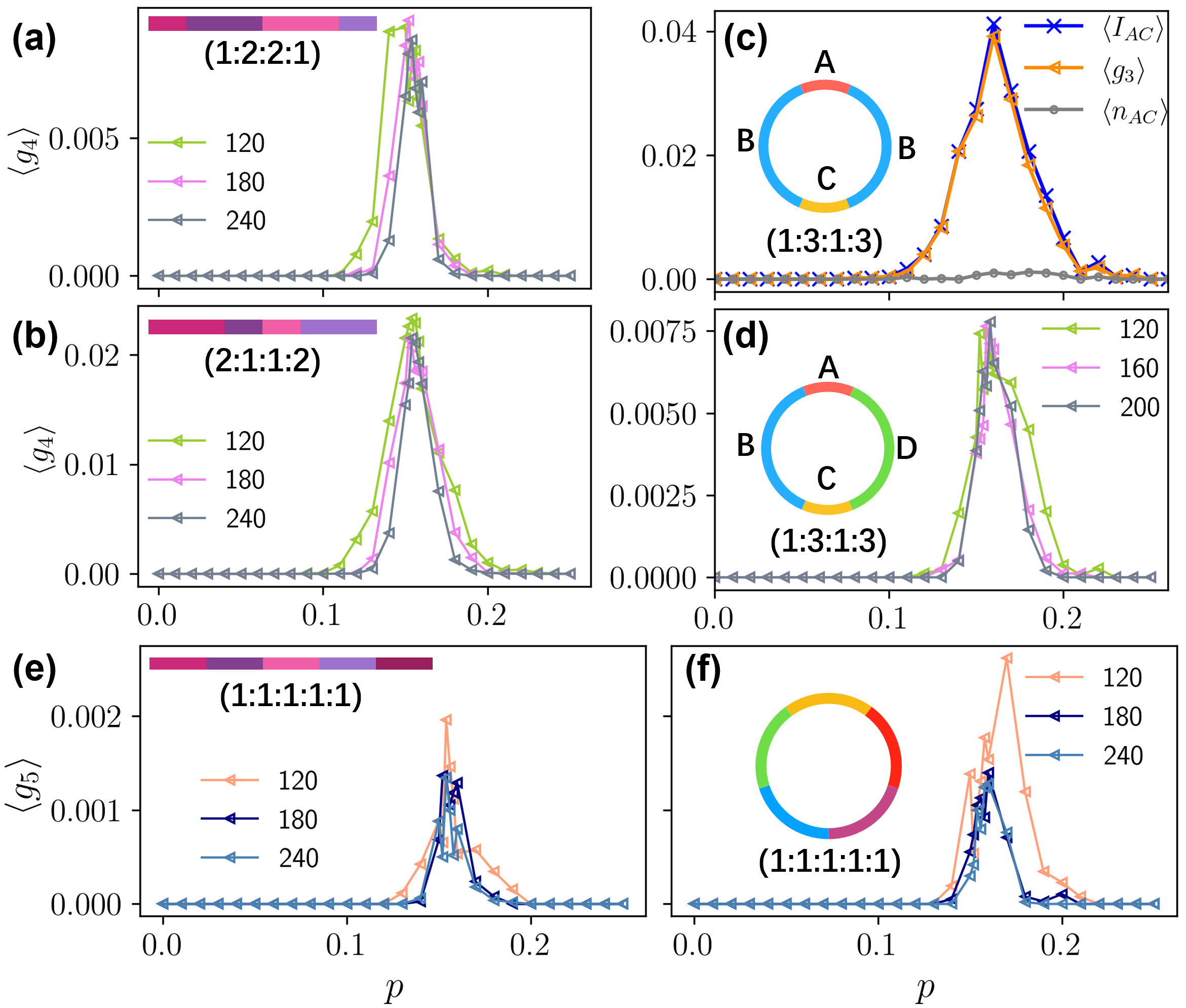} 
  \caption{$\ghz_{4,5}$ entanglement appears at 
  the measurement-induced criticality.
  Configurations are specified in the insets 
  of the subplots. (a,b,d) $\ghz_4$ and (e,f) $\ghz_5$ 
  for three different values of $N$. 
   In (c) we plot $\langle I_{AC}\rangle$, 
   $\langle n_{AC}\rangle$ and $\avgc$ for $N=120$ 
   in a tripartite setting. }\label{fig_ghz45}
\end{figure}

In this section, we move to $\ghz_n$ entanglement with $n>3$.
It might be expected that the volume-law phase is also the residence of $\ghz_{n\geq 4}$ entanglement.
However, we find that $\ghz_{n\geq 4}$ entanglement is statistically 
significant only at the measurement-induced critical point.
This is evidenced by Fig.~\ref{fig_ghz45}(a,b,d,e,f), 
where we consider a few 4 and 5-partite 
settings (sketched in each panel) and plot $\avgd$ and $\avge$, respectively. All these plots 
show peaks standing near $p_c$, similar to Fig.~\ref{fig_ghz}(b)
where $N_B/N>1/2$. In this sense, $\ghz_{n\geq 4}$
entanglement can be viewed as a measurement-induced
critical phenomenon.

\subsection{A No-Go Argument}  

Unfortunately, we do not have an analytical theory to explain what we have
numerically observed. Nevertheless, 
in some situations that will be specified below, $\ghz_{n\geq 4}$ entanglement can be 
ruled out by PIPT of $\ghz_3$ entanglement. 

The argument is based on coarse-graining of the partitioning. 
Given an $n$-partition, we can combine two neighboring parties into one and obtain 
an $(n-1)$-partition. In this sense, we can
treat $\ket{\ghz_n}$ as an $(n-1)$-partite state 
where a special party possesses two qubits. By applying a CNOT gate 
upon the two qubits belong to this party, the state
$\ket{\ghz_n}$ is transformed into $\ket{\ghz_{n-1}}\otimes |0\rangle$. 
Such CNOT gate is localized in the specified party so that the entanglement structure of
the $(n-1)$-partite state does not change. Then, we learn that
${g_n\neq 0}$ implies ${g_{{n-1}}\neq 0}$ with respect to
the coarse-grained ${(n-1)}$-partitions obtained 
from the original $n$-partition by merging
two parties into one. We can repeat this until we get a tripartition. If the original
$n$-partition has $g_n\neq 0$, we immediately know that the
tripartition has $g_3\neq 0$.

Now consider an $n$-partition with open boundary condition
where the relative size of each party, from left to right, is 
$(r_1, r_2,\dotsc, r_n)$ so that $r_i>0$ and $\sum_i  r_i=1$.
If $r_1+r_n < 1/2$[Figs.~\ref{fig_ghz45}(a) and (e) belong to
this case], there must be $g_n =0$. Otherwise, the tripartition obtained by
merging $(r_2, r_3,\dotsc, r_{n-1})$ into a single party,
which has a relative size larger than $1/2$, 
will have $g_3\neq 0$. But this result conflicts with PIPT of circuits in Config.~(1).
The same reasoning applies to the unitary circuits of Configs.~(2,3).

\subsection{Long-Range v.s. Multipartite}

In the end, we revisit a 4-partite configuration with the size ratio (1:3:1:3) illustrated in 
the inset of Fig.~\ref{fig_ghz45}(c), which has been 
studied in a pioneer work~\cite{Li:2019aa}. Therein,
the authors calculated the quantum mutual information between \textit{A} and \textit{C}, $I_{AC}$, 
and found that it is finite at $p=0.16$ but vanishes 
exponentially in $N$ elsewhere. This behavior is exactly the same with
what we have observed for $\ghz_{n\geq 4}$. 

How to understand the calculation of quantum mutual information? It is well known
that mutual information is a measure of all possible correlations whereas
entanglement is just a kind of quantum correlation.
In Ref.~\cite{Li:2019aa}, the authors claimed that $I_{AC}$ is contributed by
the bipartite \textit{A-C} entanglement. Since \textit{A} and \textit{C} are
separately, this is \emph{long-range} bipartite entanglement.
To verify this claim, we can treat \textit{B} and \textit{D} as a single party. Then,
Eq.~\eqref{eq:decomposition-3} applies so that 
it is known that $I_{AC}=g_3+2n_{AC}$, where $n_{AC}$ is  
defined in Eq.~\eqref{eq:decomposition-3}. The question is thus reduced to which one
between $g_3$ and $2n_{AC}$ that dominate the contribution to
$I_{AC}$. We show in Fig.~\ref{fig_ghz45}(c) that the answer is the former and 
actually $\langle n_{AC}\rangle\approx 0$, i.e., there is little bipartite 
long-range entanglement. It is multipartite entanglement that characterizes
the correlations at the measurement-induced criticality.


\section{Conclusions.} 
\label{sec:conclusion}

To summarize, we have revisited monitored random Clifford circuits, 
and found that (1) States in the volume-law phase 
are found to have the same 
$\ghz_3$ content, $\avgc\approx 1.25$. 
This counter-intuitive result demonstrates that $\ghz_3$ 
may capture some 
unknown structure of random circuits. (2) The dynamics of
$\ghz_3$, with respect to normalized time ($\tau$),
displays a clear stage-like profile consists of
plateaus connected by (birth and death) DPTs. 
In unitary circuits, the critical points  
of these DPTs are governed by $v_E$, the 
entanglement speed of bipartite entanglement. 
(3) $\ghz_{n\geq 4}$ is not statistically significant 
in the bulk of the volume-law phase. Instead, it is 
found at the measurement-induced 
criticality. It means that there should be a fundamental gap between 
$\ghz_{3}$ and $\ghz_{n\geq 4}$ entanglement, which
has not been noticed before.
The popular notion of ``long-range entanglement'' is not 
able to capture such gap because 
$\ghz_{3}$ and $\ghz_{n\geq 4}$ are equal with respect to  
spatial extensions.

\subsection{Discussion on DPTs of General IrME}

Among these results, we believe that the dynamical features of DPTs should be 
generalizable to IrME of broader systems. 
Indeed, in an on going work we have seen staircase 
profile of sequential 
plateaus and DPTs in the quenched dynamics of 
Markov gap (a measure of IrME that captures entanglement in the type of W state)  
in other 1D systems. 
However, it is not sure that whether the speed for IrME to build up is still $v_E$. 
But the following analysis suggests that it is strongly related to the theory of
operator growth, and more precisely, the growth of unitary generated by
localized operators.

Suppose that we have a short-range Hamiltonian. Following the 
same idea of Eq.~\eqref{em:inter-picture}, an interaction 
picture can be defined with respect to terms of the Hamiltonians that are
restricted completely within each party.
The evolution unitary in the Schr\"{o}dinger picture
reads
\be\label{em:us-uint}
U_S(t)= \big[\otimes_{i=1}^{n}U_i(t)\big] U^{\text{int}}(t)
\ee
where the first factor is irrelevant to entanglement, and 
$U^{\text{int}}(t)$ is the evolution unitary in the interaction picture
\be
U^{\text{int}}(t)=\hat{T}\exp[-i\int_{0}^t d\tau \sum_{i=1}^{n-1} 
H^{\text{int}}_{i,i+1}(\tau)].
\ee 
Therein, $\hat{T}$ is the time-ordering operator and 
$H^{\text{int}}_{i,i+1}(\tau)$ is the interaction term
between Party-$i$ and Party-($i+1$) in the 
interaction picture. At $t=0$, the operator $H^{\text{int}}_{i,i+1}$ localized at the
boarder between Party-${i}$ and Party-(${i+1}$). This is why we say
the dynamics of IrME is about the unitary generated by growing operators.

Following the idea of Lieb-Robinson bound~\cite{Lieb:1972aa}, a lower bound of the critical point can be
immediately obtained. This is because the supports of different $H^{\text{int}}_{i,i+1}(\tau)$
do not have overlaps before a certain time. Denote the Lieb-Robinson speed by
$v_{b}$~\cite{Roberts:2016aa} . Then, the time when $H^{\text{int}}_{i-1,i}(\tau)$ meets
 $H^{\text{int}}_{i,i+1}(\tau)$ will not be earlier than
\be 
t_{\min}=\frac{L}{2 v_b},
\ee 
where $L$ denotes the size of each party. Before this moment, we have
\be \label{em:u-factorize}
U^{\text{int}}(t< t_{\min})= 
\bigotimes_{i=1}^{n-1} U_{i,i+1}^{\text{int}}(t)
\ee 
where 
$U_{i,i+1}^{\text{int}}(t)=
\hat{T}\exp[-i\int_{0}^t d\tau H^{\text{int}}_{i,i+1}(\tau)]$. Such unitary resembles a
single level of our brickwork circuits, hence, is impossible to generate IrME.
For an ultimate solution to the critical time of birth-DPT of IrME, however, we still
need to understand also the content of the unitary generated by the growing operators, 
not only the spatial size of its support. This is
because these operators will not always growing---$H^{\text{int}}_{i,i+1}(\tau)$
is restricted within Parties-$i$ and $(i{+}1)$ by definition.

\begin{acknowledgments}
Thanks to Yi Zhou, Zi-Xiang Li, Run-Ze Chi, Zong-Sheng Zhou, Yan-Tao Wu, and
Yijian Zou for useful discussions. Y.-X. Zhang acknowledges the financial 
support from  Innovation Program for Quantum Science and Technology 
(Grant No. 2023ZD0301100), National Natural Science Foundation of China (Grant No. 12375024), 
CAS Project for Young Scientists in Basic Research (YSBR-100).
\end{acknowledgments}

\appendix
\section{Methods of Numerical Calculations}
\label{sec:SM_S1}
\begin{figure}[tb]
    \centering
    \includegraphics[width= 0.85\textwidth]{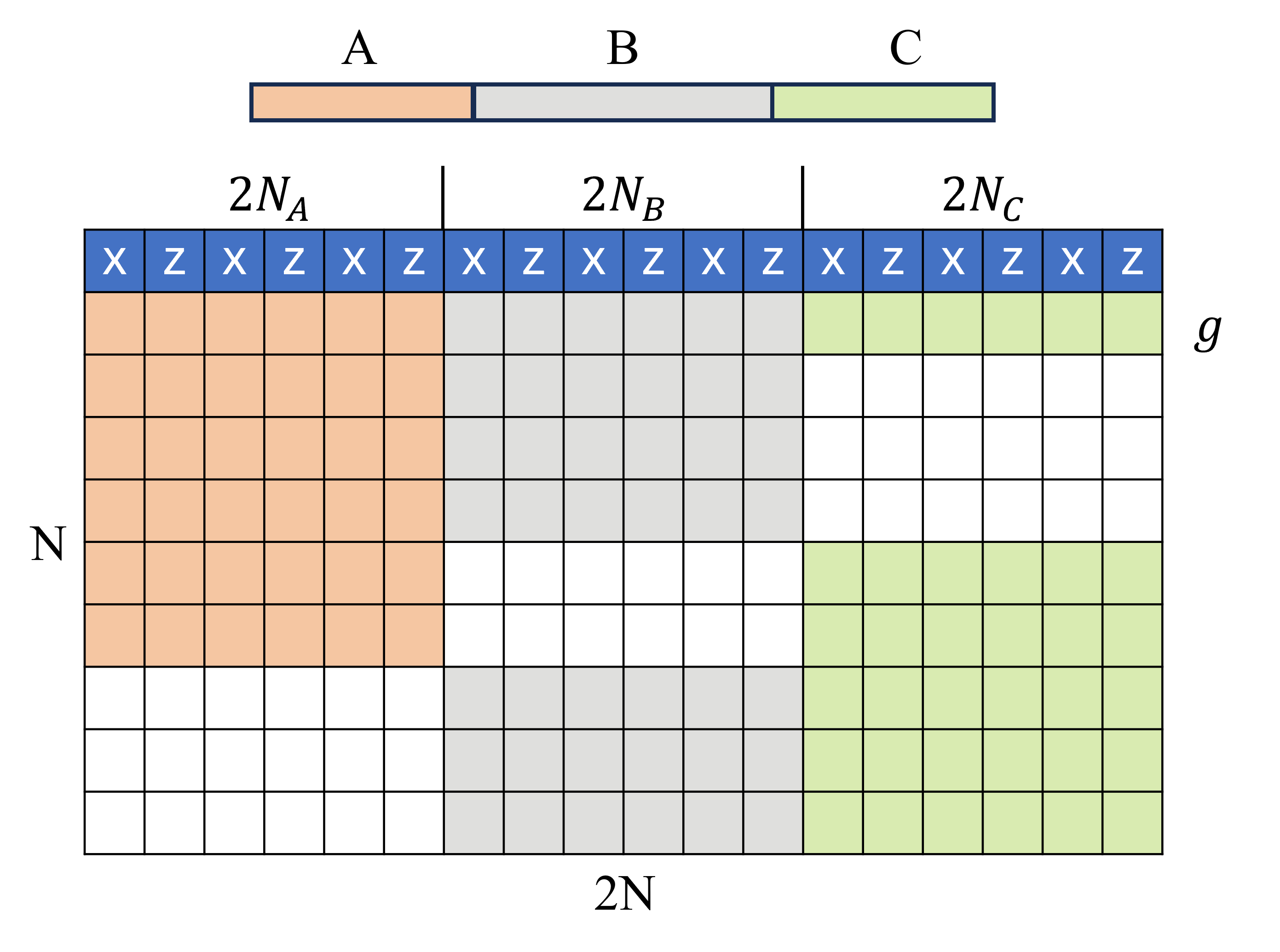}
    \caption{Sketch of a stabilizer tableau of a 9-qubit stabilizer state. The first stabilizer generator (the first row) has support on all parties, hence has
    contribution to $\ghz_3$ entanglement. The others contribute to the pairwise
    Bell entanglement.}
    \label{fig:SM1}
\end{figure}

For any intermediate state of every individual realization of the random 
Clifford circuits,
generators of the stabilizer group are
represented in the form of tableau, a table with dimension $N$-by-$2N$. 
This is because 
each stabilizer is in the form of Pauli strings. On each site, 
the Pauli words (X, Y, Z, $\mathbb{I}$) can be represented in the form of 
$X^{a}Z^b$ with $a,b\in \{0,1\}$, up to a local phase. Thus, the four Pauli operators are mapped to 2-bit strings $(ab)=(10,11,01,00)$,
respectively. During the evolution, applying one Clifford gate requires an update of all $N$ stabilizers on the relevant sites, and one layer of brickwork random Clifford gates on $N$ sites requires in total $\mathcal{O}(N^2)$ calculations. 
Further processing is needed in order to read out the GHZ index $g_n$.
We then apply Gaussian elimination on $\mathbf{Z}_2$ to each 
party and recast the tableau into the form shown in Fig.~\ref{fig:SM1}. 
Therein, the colored sections represent linear 
independent bit strings (linear independence is unnecessary if
one just needs $g_3$ but not other indices for bipartite entanglement). 
Uncolored sections are all-zero.
The GHZ index $g_3$ is equal to the number of stabilizer generators which have non-trivial support on all parties. For example, in the Fig.~\ref{fig:SM1}, $g_3=1$. 

The number of Bell states shared by Party-$i$ and Party-$j$ can be calculated as,
\begin{equation}
    n_{ij} = \frac{1}{2}\big[
    (N - \mathrm{rk}(M_k) - g_3 - \dim(S_i) - \dim(S_j) 
    \big]
\end{equation}
where $ \mathrm{rk}(M_k)$ is the rank of the sub-matrix of all columns belong to Party-$k$ ($k\neq i,j$)
and $S_i$ represents the subgroup consisting of
all stabilizers supported entirely by Party-$i$ (not shown
in Fig.~\ref{fig:SM1}). These stabilizers
are associated with qubits not contributing to entanglement.

For $g_n$ with $n>3$, we use a more direct method. 
After applying Gaussian elimination on $\mathbf{Z}_2$ to each Party-$i$, 
we record the stabilizer generators which have no support on that party, 
denoted by $\hat S_i$. Then the $\ghz_n$ index can be calculated as 
$g_n = N - \mathrm{rk}(\cup_i \hat S_i)$. That is, we just need to collect all rows of 
each $\hat{S_i}$ and count its rank. 
Our code is developed on Julia, and using the QuantumCliffod package.

\section{Odd-Even effect}
\label{app:odd-even}
\begin{figure}[b]
    \centering
    \includegraphics[width=0.85\textwidth]{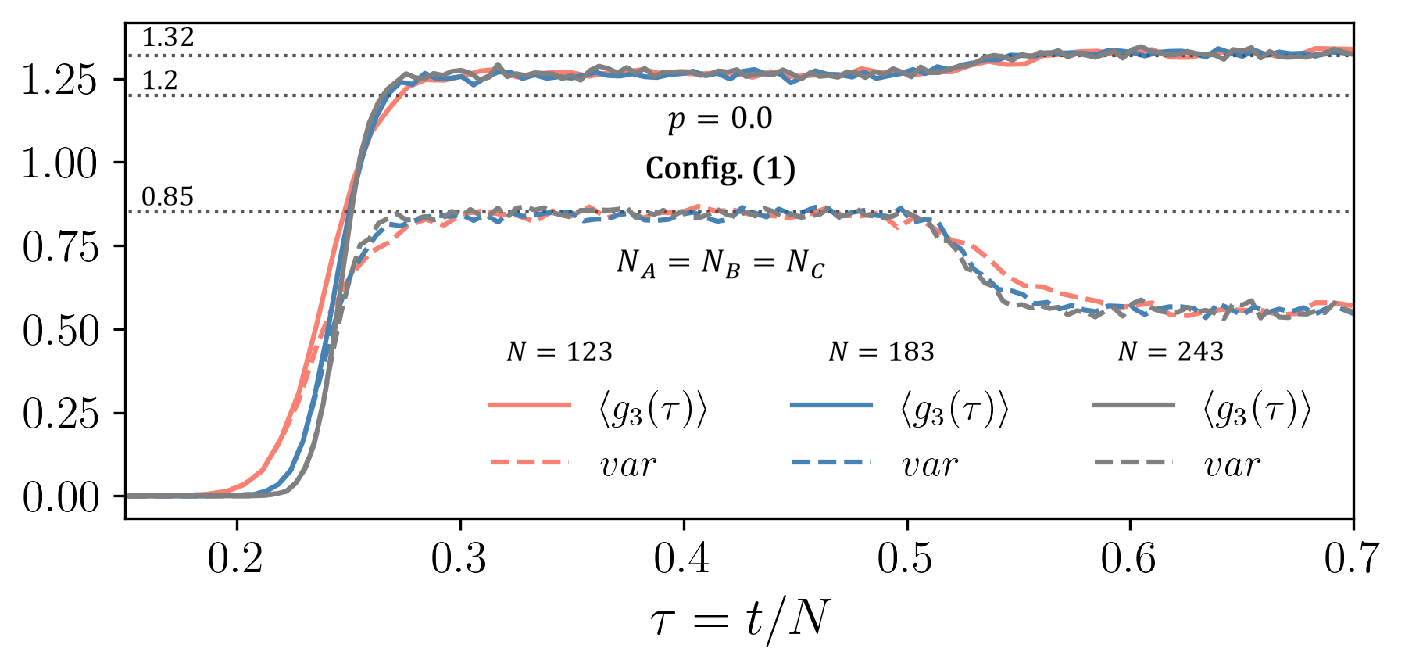}
    \caption{
        Counterpart of Fig.~\ref{fig_transient}(a) but with 
        odd number of qubits $N=123, 183, 243$.}
    \label{fig:SM_odd}
\end{figure}

We have seen a slight lower plateau of $\avgc\approx 1.2$ in circuits with $p=0$. This is actually 
an odd-even effect. If $N$ is an odd number, a higher plateau is 
obtained instead. This is illustrated in Fig.~\ref{fig:SM_odd}
for an equal partitioning with $N=123, 183, 243$.
Therein, we see another 3-stage profile there the 3rd plateau
is about $\avgc\approx 1.32$. Meanwhile, the variance becomes
smaller than the case of even $N$.

\section{$\ghz_3$ calculation of N-qubit random Clifford circuit}
\label{app:SM_S3}

\begin{figure}[tbh]
	\centering
	\includegraphics[width = 0.8\linewidth]{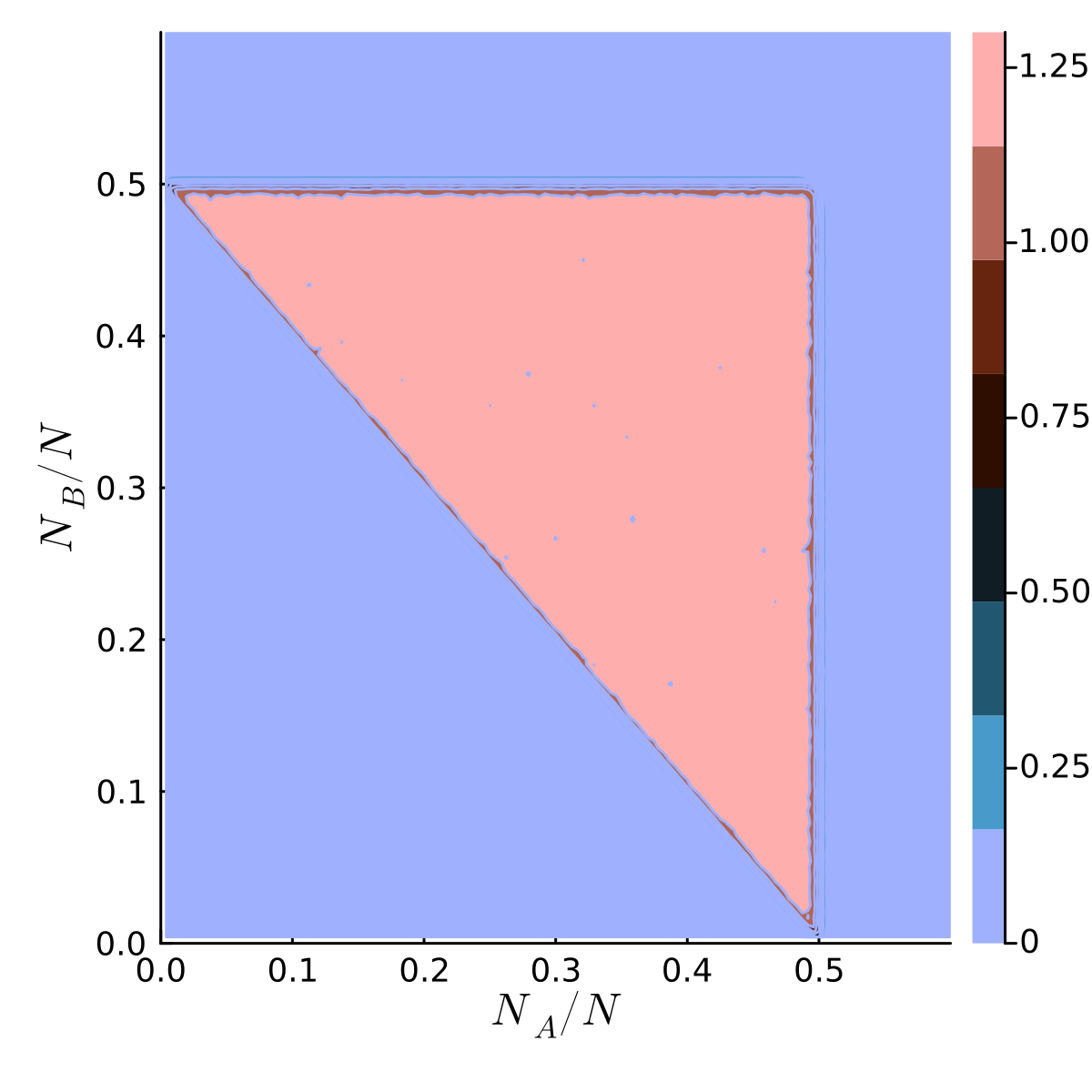}
	\caption{$\ghz_3$ of random N-qubit Clifford circuits with respect to different partitions.}
	\label{fig:SM_triangle}
\end{figure}

Since brickwork circuits of linear depth approximately sample the N-qubit 
Clifford circuit~\cite{Aaronson:2004aa, Maslov:2018, Bravyi:2021, Moore:2001, Jiang:2020}, 
we focus on the $\langle g_3 \rangle$ of states 
generated by a random N-qubit Clifford gate. 
Firstly, we generate random N-qubit Clifford gate using Julia package `QuantumClifford.jl' and scan 
$\ghz_3$ over all possible tripartite configurations ($N_A, N_B, N_C$) with the total number $N = 240$. 
The results are shown in Fig.~\ref{fig:SM_triangle}. 
The $\avgc$ is obtained by average over 2000 realizations.
In this figure, there are three sharp boundaries ($N_A/N = 0.5, N_B/N = 0.5, (N_A + N_B)/N = 0.5$) 
between the non-zero and zero $\ghz_3$ regimes. These three conditions are corresponding to `Triangle Inequality Theorem'. 

Next, we use the following equality~\cite{Nezami:2020aa} to obtain an upper bound for $\avgc$:
\begin{equation}
    g_3 = \log_2( \Tr[ (\rho_{AB}^{T_B}) ^3 ] ) + S(A) + S(B) + S(C),
\end{equation}
where ${T_B}$ denotes the partial transpose on the subsystem B,
$S(A)$ represents the von Neumann entropy of
subsystem A and similarly to $S(B)$ and $S(C)$. The equality can be directly 
obtained from Eq.~\eqref{eq:decomposition-3}.
We express the dependence of $\rho$ on $U$ implicitly, so that
the expectation value of $g$ averaged over the N-qubit Clifford group
reads
\begin{align}
    \label{eq:g_avg}
    \langle g_3 \rangle = 
    \frac{1}{\mathcal{N}}\sum_U  \bigg(& \log_2( \Tr[ (\rho_{AB}^{T_B}) ^3 ] ) \notag \\
    &+ S(A) + S(B) + S(C) \bigg),
\end{align}
where $\mathcal{N}$ denotes the size of the N-qubit Clifford group.
For the logarithmic term, the Jensen inequality of concave function $\log_2(x)$ indicates that 
\begin{align}
    \label{eq:haar_random_ineq}
    \frac{1}{\mathcal{N}} \sum_U   \log_2( &\Tr[ (\rho_{AB}^{TB}) ^3 ] ) \notag \\
    &\leq \log_2 \left(  \frac{1}{\mathcal{N}} \sum_U  \Tr[ (\rho_{AB}^{T_B}) ^3 ] \right).
\end{align}
To linearize the formula, we apply the triple replica trick 
through two permutation operators $R_A^{+}$ and $R_B^{-}$ acting on three copies of $\rho_{A}$
and $\rho_B$, respectively,
\begin{align}
     &\frac{1}{\mathcal{N}} \sum_U  \Tr[ (\rho_{AB}^{T_B}) ^3 ] = \frac{1}{\mathcal{N}}\sum_U \Tr[ \rho_{AB}^{\otimes 3} R_A^+ R_B^{-} ] \notag \\ 
     &=  \Tr[ \left( \frac{1}{\mathcal{N}}\sum_U  \rho^{\otimes 3} \right) R_A^{+} R_B^{-} ]
     \label{eq:trace_replica}
\end{align}
Operator $R_A^{+}$ permutes the Hilbert spaces of the three copies 
according to $\ket{k}_1 \to \ket{k}_2 \to \ket{k}_3 \to \ket{k}_1$ 
for every basis state $\ket{k}$. $R_B^{-}$ works similarly but
on the opposite direction
$\ket{k}_1 \to \ket{k}_3 \to \ket{k}_2 \to \ket{k}_1$. 

As the multiqubit stabilizer states are projective 3-designs\cite{Zhu:2017,Webb:2015vwz}, The average of the states can be written as the summation of permutation operators with equal coefficients,
\begin{align}
     \frac{1}{\mathcal{N}}\sum_U   \rho^{\otimes 3} &= \frac{1}{\alpha} \sum_{\pi=0} ^5 \rho_\pi \\
     \rho_0 &= \mathbf{1} \\
     \rho_1 &= \textit{\rm SWAP}_{12} \otimes \mathbf{1}_3 \\
     \rho_2 &= \textit{\rm SWAP}_{23} \otimes \mathbf{1}_1 \\
     \rho_3 &= \textit{\rm SWAP}_{13} \otimes \mathbf{1}_2 \\
     \rho_4 &= \rho_1 \rho_2 \\
     \rho_5 &= \rho_2 \rho_1 
\end{align}
where $\alpha = d(d+1)(d+2)$, $d =2^N$ and $N$ is the system size. 
The trace of permutation operators $\Tr{\rho_0}=2^{3*N}$, $\Tr{\rho_1}=\Tr{\rho_2}=\Tr{\rho_3}=2^{2*N}$, $\Tr{\rho_4} = \Tr{\rho_5} = 2^{N}$. A straightforward insight at these results is that the trace of a general "SWAP" gate is the number of invariant states. So, SWAP between two copies has the invariant states $\ket{\phi_\mu}_1 \otimes \ket{\phi_\mu}_2$, where $\mu \in [1,2^{d_\phi}]$. $\Tr{\text{\rm SWAP}_{12}} = 2^N$. 
The general SWAP between three copies shares the same nature while invariant states are $\ket{\phi_\mu}_1 \otimes \ket{\phi_\mu}_2 \otimes \ket{\phi_\mu}_3$. The subscriptions of $\text{\rm SWAP}$ represent the copies of the density matrix. 

In this notation, the permutations $R_A^+$ and $R_B^-$  of subsystem (A,B,C) among different copies (1,2,3) can be written as,
\begin{align}
    R_A(+) &= (\text{\rm SWAP}_{12}^A \otimes \mathbf{1}_3^A) (\mathbf{1}_1^A \otimes \text{\rm SWAP}_{23}^A ) \otimes \mathbf{1}_{123}^B \otimes \mathbf{1}_{123}^C \\
    R_B(-) &= \mathbf{1}_{123}^A \otimes (\mathbf{1}_1^B \otimes \text{\rm SWAP}_{23}^B )(\text{\rm SWAP}_{12}^B \otimes \mathbf{1}_3^B) \otimes  \mathbf{1}_{123}^C
\end{align}
For the trace of each terms of $\rho_\pi R_A(+) R_B(-)$, 
\begin{align}
    &\Tr[\rho_0  R_A(+) R_B(-) ] = \Tr[ R_A(+) R_B(-)] \notag \\
    &= \Tr \bigg( (\text{\rm SWAP}_{12}^A \otimes \mathbf{1}_3^A) (\mathbf{1}_1^A \otimes \text{\rm SWAP}_{23}^A )  \notag \\
    &\otimes (\mathbf{1}_1^B \otimes \text{\rm SWAP}_{23}^B )(\text{\rm SWAP}_{12}^B \otimes \mathbf{1}_3^B) \mathbf{1}_{123}^C \bigg) \notag \\
    &= 2^{N_A} \cdot 2^{N_B} \cdot 2^{3N_C} = 2^{N +2N_c} 
\end{align}
Similarly, we obtain that
\begin{align}
    &\Tr{\rho_1  R_A(+) R_B(-) } = 2^{2N_A} \cdot 2^{2N_B} \cdot 2^{2N_C} = 2^{2 N} \\
    &\Tr{\rho_2  R_A(+) R_B(-) } = 2^{2N_A} \cdot 2^{2N_B} \cdot 2^{2N_C} = 2^{2 N} \\
    &\Tr{\rho_3  R_A(+) R_B(-) } = 2^{2N_A} \cdot 2^{2N_B} \cdot 2^{2N_C} = 2^{2 N} \\
    &\Tr{\rho_4  R_A(+) R_B(-) } = 2^{N_A} \cdot 2^{3N_B} \cdot 2^{N_C} = 2^{N+2N_B} \\
    &\Tr{\rho_5  R_A(+) R_B(-) } = 2^{3N_A} \cdot 2^{N_B} \cdot 2^{N_C} =2^{N + 2N_A} 
\end{align}
So the averaged value inside of the logarithmic braket of the upper bound of $\langle g \rangle$, i.e, Eq.(\ref{eq:trace_replica}), is
\begin{align}
    Eq.(\ref{eq:trace_replica}) &= \frac{1}{\alpha} \left( 2^N(2^{2N_A} + 2^{2N_B} + 2^{2N_C} ) + 3*2^{2N} \right) \notag \\
    &= \frac{3*2^{2N} + 2^N(2^{2N_A} + 2^{2N_B} + 2^{2N_C} )}{(2^N)(2^N+1)(2^N+2)} \notag \\
    & < \frac{3}{2^N} + \frac{2^{2N_A} + 2^{2N_B} + 2^{2N_C} }{(2^N+1)(2^N+2)} 
\end{align}
For an equally tripartite system ($N_A = N_B = N_C = 1/3N$) the upper bound of the logarithmic term in the thermal limit is then:
\begin{align}
    \lim_{N >> 1}  \frac{3}{2^N} + \frac{2^{2N_A} + 2^{2N_B} + 2^{2N_C} }{(2^N+1)(2^N+2)} = \frac{3}{2^N}
\end{align}
Together with Eq.~\eqref{eq:g_avg} and Eq.~\eqref{eq:haar_random_ineq}, we obtain an upper bound of $\langle g \rangle$,
\begin{align}
    \langle g_3 \rangle &< \log_2(\frac{3}{2^N}) + S(A) + S(B) + S(C) \notag \\
    & = \log_2(3) - N  + S(A) + S(B) + S(C) \notag \\
    & < \log_2(3)
\end{align}

For a system partition that has one partition is more than half of the system ($ N_A \ge 1/2N > N_B \ge N_C$) the upper bound of the logarithmic term in the thermal limit is then:
\begin{align}
    \lim_{N >> 1}  \frac{3}{2^N} + \frac{2^{2N_A} + 2^{2N_B} + 2^{2N_C} }{(2^N+1)(2^N+2)} = \frac{2^{2N_A}}{2^{2N}}
\end{align}
Together with Eq.~\eqref{eq:g_avg} and Eq.~\eqref{eq:haar_random_ineq}, we obtain an upper bound of $\langle g \rangle$,
\begin{align}
    \langle g_3 \rangle & \leq \log_2(\frac{2^{2N_A}}{2^{2N}}) + S(A) + S(B) + S(C) \notag \\
    & \leq 2N_A - 2N  + (N_B + N_C) + N_B + N_C \notag \\
    & =0
\end{align}
In this case, $\langle g_3 \rangle = 0$.

\section{$\ghz_3$ Entanglement in circuits of Configs.~(2) and (3)}
\label{app:config23}
\begin{figure}[htb!]
	\centering
	\includegraphics[width = \linewidth]{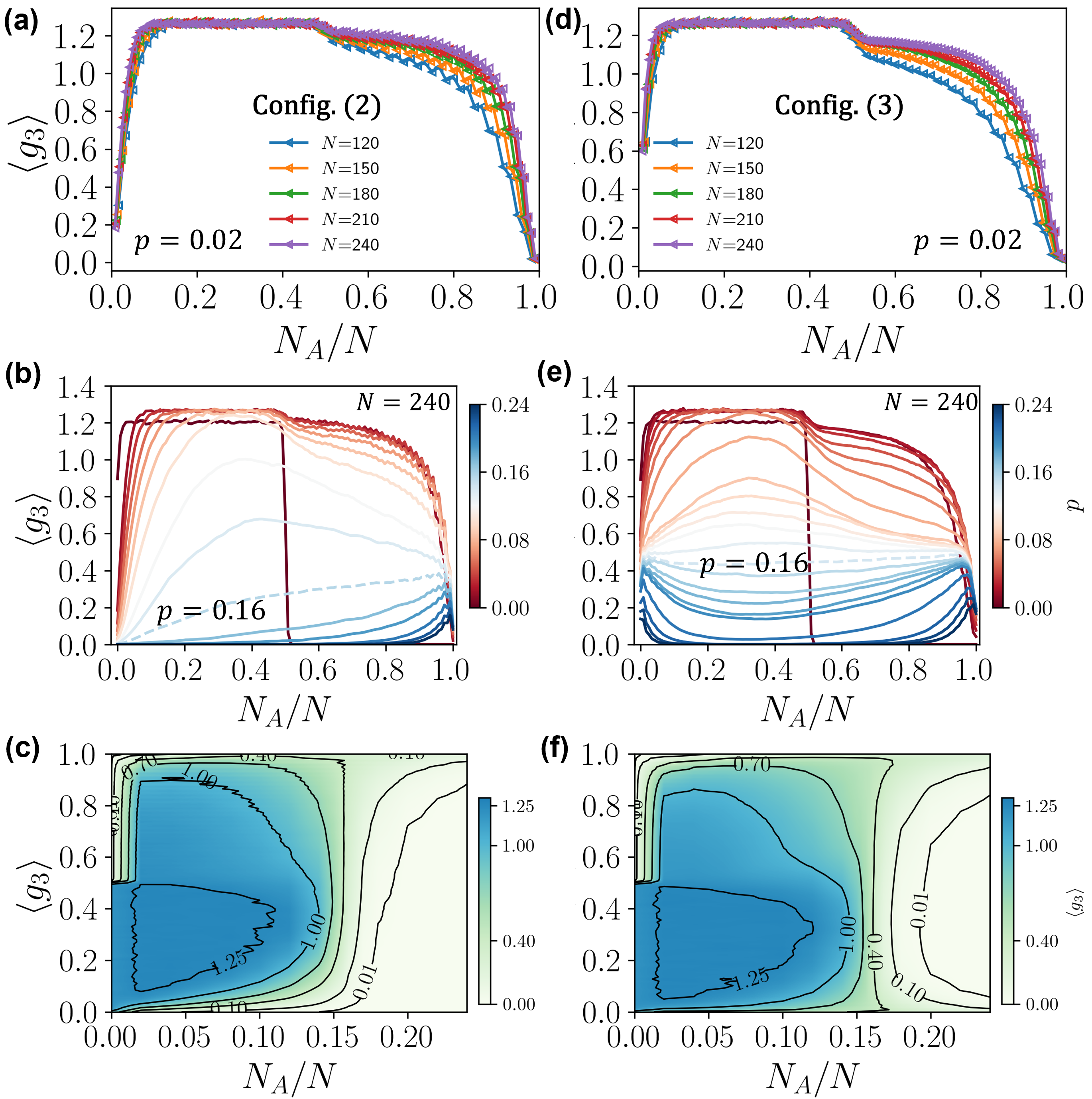}
	\caption{$\avgc$ of Config.~(2) (a, b, c) and Config.~(3) (d, e, f)  with respect to $N_A / N$. (a, d) $\avgc$ reaches a plateau as the $N_A/N$ increases and split into $N$-dependent declines when $N_A/N \ge 1/2$. (b, e) $\avgc$ as a function of $N_A/N$ for $N=240$ with different $p$. (c, f) $\avgc$ as a function of $p$ and $N_A/N$ for $N=240$.}
	\label{fig:SM2_NA}
\end{figure}

In Fig.~\ref{fig_ghz}  we have elaborated on the partitioning-induced phase transition (PIPT) of $\ghz_3$ 
entanglement in circuits of Config.~(1). Here we present 
the results of the other two configurations: Config.~(2) with
open boundary condition and Config.~(3) with periodic boundary
condition.

\subsection{Config.~(2)}

In Fig.~\ref{fig:SM2_NA}(a), we fix the measurement probability
at $p = 0.02$ and plot $\avgc$ as a function of partitions parameterized
by $N_A/N$ for five different values of $N$. Comparing it with
Figs.~\ref{fig_ghz}(e) and (f), we immediately see that there
is no PIPT because $\avgc$ does not drop to
zero when $N_A/N>1/2$. However, the plateau at $\avgc\approx 1.25$,
which is common for all plotted values of $N$, splits into
$N$-dependent declines when $N_A/N$ exceeds $1/2$. A convergence of
such declines may appear if we further
increase the value of $N$. 

Nevertheless, the appearance of significant sizing effect
marks something unusual at $N_A/N=1/2$.
Recall that in the case of $N_A/N>1/2$, there is no steady-state 
$\ghz_3$ entanglement if $p=0$, due to the sudden death. 
Thus, the finite $\ghz_3$ entanglement seen in
Fig.~\ref{fig:SM2_NA}(a) with $p>0$ should be attributed to 
the effect of measurements, and $N_A/N=1/2$ is the onset of such effect.

Then we fix $N=240$ but consider 
various values of $p$ in the range of $0\leq p\leq 0.24$
in Fig.~\ref{fig:SM2_NA}(b), which is
the counterpart of Fig.~\ref{fig_ghz}(d) for Config.~(1).
It shows two distinguished features: $\avgc$ drops to zero at $N_A/N=1/2$ when
$p=0$ (dark red curve), and the curve of $p=0.16$ (dashed light blue
curve) is roughly linear in $N_A/N$. 
Curves of $p<0.16$ are concave in $N_A/N$ and curves of $p>0.16$ are
convex, showing that $p=0.16$ is a transition point.
Moreover, the plateau of
$p=0$ is slightly lower than $p\neq 0$, which is same with
the case of Config.~(1).

We plot the counterpart of Fig.~\ref{fig_ghz}(c) in Fig.~\ref{fig:SM2_NA}(c). 
It looks sharply different from
Fig.~\ref{fig_ghz}(c) because PIPT is suppressed to only the
point of $p=0$. We see quick changes of $\avgc$ 
near the line of $p=0$ and $N_A/N\geq 1/2$.

\subsection{Config.~(3)}
We show the parallel plots of $\ghz_3$ entanglement
of Config.~(3) in Fig.~\ref{fig:SM2_NA}. Compared
with the case of Config.~(2), the most conspicuous
difference is found in Fig.~\ref{fig:SM2_NA}(e). Therein,
we see a congruence of all curves at $\avgc\approx 0.5$
for small $N_A/N$, and the curve of $p=0.16$ 
(dashed light blue curve) is roughly flat, indicating
scale invariance of $\avgc$ at the measurement-induced criticality.

\bibliography{randomGHZ}

\end{document}